\documentclass[twoside]{article}

\usepackage[accepted]{aistats2022}

\usepackage[round]{natbib}

\usepackage[british]{babel}

\usepackage{amssymb}
\usepackage{mathtools} 
\usepackage{amsthm}
\usepackage{centernot}
\usepackage{thmtools}
\usepackage{thm-restate}

\usepackage{graphicx}
\usepackage{wrapfig}
\usepackage{subcaption}
\usepackage{booktabs} 
\graphicspath{{./figures/}}
\usepackage{adjustbox}
\usepackage{caption}
\usepackage{tikz}
\usetikzlibrary{shapes.geometric,positioning,bayesnet,calc}
\usetikzlibrary{arrows,shapes,plotmarks,positioning}
\usetikzlibrary{decorations.markings}
\tikzset{>=stealth'} 
\tikzstyle{graphnode} = 
   [circle,draw=black,minimum size=22pt,text centered,text
     width=22pt,inner sep=0pt] 
\tikzstyle{var}   =[graphnode,fill=white]
\tikzstyle{vardashed}   =[graphnode,draw=gray,fill=white]
\tikzstyle{obs}   =[graphnode,fill=black,text=white]
\tikzstyle{obsgrey}   =[graphnode,draw=white,fill=lightgray,text=black]
\tikzstyle{par}    =[graphnode,draw=white,fill=red,text=black] 
 \tikzstyle{crucial} =[graphnode,draw=white,fill=yellow,text=black] 
\tikzstyle{fac}   =[rectangle,draw=black,fill=black!25,minimum size=5pt]
\tikzstyle{facprior} =[rectangle,draw=black,fill=black,text=white,minimum size=5pt]
\tikzstyle{edge}  =[draw=white,double=black,very thick,-]
\tikzstyle{blueedge}  =[draw=white,double=blue,very thick,-]
\tikzstyle{rededge}  =[draw=white,double=red,very thick,-]
\tikzstyle{prior} =[rectangle, draw=black, fill=black, minimum size=
5pt, inner sep=0pt]
\tikzstyle{dirprior} = [circle, draw=black, fill=black, minimum
size=5pt, inner sep=0pt]
\tikzstyle{dot_node}=[draw=black,fill=black,shape=circle]

\usepackage{hyperref}
\usepackage{url}
\usepackage{cleveref}

\crefname{section}{sec.}{secs.}
\Crefname{section}{Section}{Sections}

\usepackage{enumitem}
\usepackage{csquotes}
\usepackage{pifont}

\newcommand{\Exp}[2]{\mathbb{E}_{#1}\left[#2\right]}

\newcommand{\CI}{\mathrel{\perp\mspace{-10mu}\perp}}
\newcommand{\nCI}{\centernot{\CI}}

\newcommand{\bx}{{\bf x}}
\newcommand{\bbx}{\bar{\bx}}

\newcommand{\bz}{{\bf z}}
\newcommand{\bX}{{\bf X}}
\newcommand{\bbX}{\bar{\bX}}

\newcommand{\bZ}{{\bf Z}}

\newcommand{\R}{{\mathbb R}}
\newcommand{\N}{{\mathbb N}}

\newcommand{\cB}{{\cal B}}
\newcommand{\cC}{{\cal C}}

\newcommand{\cN}{{\cal N}}

\newcommand{\cX}{{\cal X}}

\newcommand{\bsh}{\backslash}

\newtheorem{Theorem}{Theorem}
\newtheorem{Lemma}{Lemma}
\newtheorem{Definition}{Definition}

\makeatletter
\def\blfootnote{\xdef\@thefnmark{}\@footnotetext}
\makeatother

\begin{document}
\twocolumn[

\aistatstitle{Obtaining Causal Information by Merging Datasets with MAXENT}

\aistatsauthor{Sergio~Hernan~Garrido~Mejia$^{*\dagger}$
\And
Elke~Kirschbaum$^{*}$
\And
Dominik~Janzing
}

\aistatsaddress{
Danmarks Tekniske Universitet\\ Lyngby, Denmark \\
\texttt{sermej@amazon.com}
\And
Amazon Research\\ Tübingen, Germany\\
\texttt{elkeki@amazon.com} \\
\And
Amazon Research\\ Tübingen, Germany\\
\texttt{janzind@amazon.com}
}
]
\blfootnote{*Both authors contributed equally}
\blfootnote{$\dagger$This work was done while the author worked at Amazon Research}
\setcounter{footnote}{0}

\runningauthor{Garrido~Mejia, Kirschbaum, Janzing}

\begin{abstract}
The investigation of the question ``which treatment has a causal effect on a target variable?'' is of particular relevance in a large number of scientific disciplines. This challenging task becomes even more difficult if not all treatment variables were or even cannot be observed jointly with the target variable. In this paper, we discuss how causal knowledge can be obtained without having observed all variables jointly, but by merging the statistical information from different datasets. We show how the maximum entropy principle can be used to identify edges among random variables when assuming causal sufficiency and an extended version of faithfulness, and when only subsets of the variables have been observed jointly. 
\end{abstract}


\section{INTRODUCTION}

The scientific community is rich in observational and experimental studies that consider a tremendous amount of problems from an even more significant number of perspectives. All these studies have collected data containing valuable information to investigate the research question at hand. At the same time, it is often impossible to use the collected data to answer slightly different or more general questions, as the required information cannot be extracted from the already existing datasets. 

Consider, for instance, a case in which we want to investigate the influence of the place of residence on the probability to become depressed, and we are given four different studies: (1) showing the depression rates for different regions; 
(2) capturing depression rate with respect to (w.r.t.) age; 
(3) providing information about the depression rate w.r.t.\ sex; and 
(4) showing the distribution of age and sex across different regions. 
We want to know whether there is a direct causal link between the place of residence and the depression rate or only an indirect link through age and/or sex. In this paper, we address the question of how we can obtain this causal information without performing a new study in which we observe all factors (age, sex, place of residence, and depression rate) at the same time, but only by merging the already collected datasets. 

Since the problem of inferring the joint distribution from a set of marginals is heavily underdetermined \citep{Kellerer1964}, we use the maximum entropy (MAXENT) principle to infer the joint distribution that maximises the joint entropy subject to the observed marginals. 
This has the advantage that the MAXENT distribution contains some information about the existence of causal arrows that also hold for the true joint distribution regardless of how much the MAXENT distribution deviates from the latter.  

As usual, our causal conclusions require debatable assumptions that link statistical properties of distributions from passive observations to causality. Therefore we use assumptions common in causal discovery \citep{Spirtes1993,Pearl2000}. Additionally, we define and intuitively justify the notion of {\it faithful $f$-expectations}, which is analogous to faithfulness in the sense of postulating genericity of parameters. This allows us to draw the following conclusions, which are the main contributions of this paper:
\begin{itemize}
\item The presence or absence of direct causal links can be identified only from the Lagrange multipliers of the MAXENT solution if the causal order is known (see \cref{sec:merging_datasets}, \cref{lm:no_edge}). 
\item For a causal graph $G$ with $N$ nodes for which the given constraints define all bivariate distributions uniquely, the graph constructed from the MAXENT distribution by connecting two nodes if and only if there is a non-zero Lagrange multiplier corresponding to some bivariate function of the two variables, is a supergraph of the moral graph of $G$ (see \cref{sec:merging_datasets}, \cref{thm:moral}). 
\item Merging datasets with MAXENT improves the predictive power compared to using the observed marginal distributions (see \cref{sec:merging_datasets}, \cref{th:predictive_power}).
\end{itemize}

The remainder of this paper is structured as follows: We start by presenting the notation and assumptions used throughout this paper. Then, in \cref{sec:maxent} we introduce the MAXENT principle. 
In \cref{sec:merging_datasets} we discuss how we can obtain causal information by merging datasets.
In \cref{sec:related_work} we put our work into the context of the related literature. Finally, in \cref{sec:experiments} we evaluate the identification of causal edges from MAXENT on simulated and real-world datasets.

{\bf Notation}\label{sec:notation}
Let $\bX=\left\{X_1,\dots,X_N\right\}$ be a set of discrete random variables. Although the results in this article hold also for continuous variables with strictly positive densities ($p(\bx)>0$) and finite differentiable entropy, for notational convenience we consider discrete random variables with values $\bx\in\cX$. Further let $X_i,X_j\in\bX$ be two variables whose causal relationship we want to investigate. We denote with $\bZ=\bX\bsh\left\{X_i,X_j\right\}$ the complement of $\left\{X_i,X_j\right\}$ in $\bX$, where (by slightly overloading notation) bold variables represent sets and vectors of variables at the same time. 
We consider the set of functions $f=\left\{f_k\right\}$ with $f_k:\cX_{S_k}\to\R$ for some $k\in\N$ and $\bX_{S_k}\subseteq\bX$. The empirical means of $f$ for a finite sample from the joint distribution $P(\bX)$ are collected in the set $\tilde{f}=\left\{\right.$$\tilde{f}_k$$\left.\right\}$, and the set of true expectations we denote with $\Exp{p}{f}=\left\{\sum_\bx p(\bx)f_k(\bx_{S_k})\right\}$. 
Further, we denote with $P$ the \enquote{true} joint distribution of the variables under consideration and with $\hat{P}$ the approximate MAXENT solution satisfying the constraints imposed by the expectations of $f$, as described in \cref{sec:maxent}. 

{\bf Assumptions}
If not stated differently, we make the following assumptions throughout this paper: The set of variables $\bX$ is causally sufficient, that is, there is no hidden common cause $U \notin \bX$ that is causing more than one variable in $\bX$ (and the causing paths go only through nodes that are not in $\bX$)\citep{peters2017elements}. Furthermore, their joint distribution $P(\bX)$ satisfies the causal Markov condition and faithfulness w.r.t.\ a directed acyclic graph (DAG) $G$ (see \cref{sec:graphical_models}).  We have $L$ datasets, where each contains observations for only a subset of the variables, and at least one dataset contains observations for the set $\left\{X_i,X_j\right\}$. The observations are drawn from the same joint distribution $P(\bX)$.\footnote{In \cref{app:different_contexts} we sketch the case where each dataset is from a different joint distribution by introducing an additional variable for the background conditions.} Further, the set of functions $f$ is linearly independent.


\section{MAXIMUM ENTROPY}\label{sec:maxent}
The maximum entropy (MAXENT) principle \citep{jaynes1957information} is a framework to find a \enquote{good guess} for the distribution of a system if only a set of expectations for some feature functions $f$ is given. The MAXENT distribution is the solution to the optimisation problem
\begin{align}\label{eq:maxent_op_eq}
&\max_{p} H_p(\bX) \notag\\
\qquad \text{s.t.: }\; &\Exp{p}{f} = \tilde{f} \; ,\quad \sum_\bx p(\bx) = 1 \; ,
\end{align} 
for the Shannon entropy $H_p(\bX)=-\sum_\bx p(\bx)\log p(\bx)$. 
Often the statistical moments $f_k(x)=x^k$ for $k\in\mathbb{N}$ are used. Note that many quantities of interest are simple expressions from expectations of appropriate functions, e.g.\ the covariance of two random variables is $\mathbb{E}\left[X_iX_j\right] - \mathbb{E}\left[X_i\right]\mathbb{E}\left[X_j\right]$. 

{\bf Approximate MAXENT}
The empirical means of the functions $f$ gained from a finite sample will never be {\it exactly} identical to the true expectations. This implies that not even the true distribution necessarily satisfies the constraints imposed on the MAXENT distribution. This can lead to large (or even diverging) values of the parameters which overfit statistical fluctuations. To account for this, the expectations only need to be {\it close} to the given empirical means. This leads to the formulation of {\it approximate} MAXENT \citep{dudik2004performance,altun2006unifying}, where the constraints in the optimisation problem in \cref{eq:maxent_op_eq} are replaced by approximate constraints, resulting in \begin{align}\label{eq:maxent_op}
&\min_{p} -H_p(\bX) \notag\\
\qquad \text{ s.t.:}\; &\|\mathbb{E}_{p}\left[f\right] - \tilde{f}\|_\cB \leq \varepsilon \; ,\quad  \sum_\bx p(\bx) = 1 \; ,
\end{align} 
with $\varepsilon\geq 0$. This type of convex optimisation problems have been studied for infinite dimensional Banach spaces $\cX$ and $\cB$ by \citet{altun2006unifying}, where it was shown that \cref{eq:maxent_op} is equivalent to
\begin{align}\label{eq:maxent_op_dual}
\max_\phi \left<\phi,\tilde{f}\right> - \log\sum_\bx\exp\left[\left<\phi,f\right>\right] - \varepsilon\|\phi\|_{\cB^*} \; ,
\end{align}
with $\cB^*$ being the dual to $\cB$. 
In contrast to standard MAXENT, whose well-known dual is maximum likelihood estimation, in approximate MAXENT, the parameters $\phi$ are regularised depending on the choice of the norm in \cref{eq:maxent_op}. For instance, $\cB=\ell_\infty$ results in a Laplace regularisation $\varepsilon\|\phi\|_1$.
Appropriate choices for $\varepsilon$ are proportional to $\mathcal{O}(1/\sqrt{M})$, where $M$ is the sample size, although in practice $\varepsilon$ is usually chosen using cross-validation techniques \citep{dudik2004performance,altun2006unifying}. 

Throughout this paper, we will use approximate MAXENT and assume that $\cX$ and $\cB$ are finite-dimensional. We consider the $\ell_\infty$ norm, which results in the element-wise constraints
\begin{align}\label{eq:maxent_constraints}
|\mathbb{E}_{p}\left[f_k\right] - \tilde{f}_k| \leq \varepsilon_k \quad \forall k 
\end{align}
for $\varepsilon_k\geq 0$. 
In this case, the MAXENT optimisation problem can be solved analytically using the Lagrangian formalism of constrained optimisation and the solution reads
\begin{equation}\label{eq:maxent_sol}
\hat{p}(\bx) = \exp\left[\sum_{k} \lambda_{k} f_{k}(\bx_{S_k}) -\alpha\right] \;,  
\end{equation}
where $\lambda=\left\{\lambda_{k}\right\}$ are the Lagrange multipliers and $\alpha=\log\sum_\bx\exp\left[\sum_k  \lambda_{k} f_{k}(\bx_{S_k})\right]$ is the partition function ensuring that $\hat{p}$ is correctly normalised. 
The optimal Lagrange multipliers can be found via
\begin{align}\label{eq:maxent_lambdas}
\min_\lambda &-\sum_k\lambda_k\tilde{f}_k + \log\sum_\bx\exp\left[\sum_k\lambda_kf_k(\bx_{S_k})\right] \notag\\
&+ \sum_k\varepsilon_k|\lambda_k| \; .
\end{align}

{\bf Conditional MAXENT}
In cases where the marginal distribution of a subset of the variables is already known, the MAXENT approach can be natively extended to a {\it conditional} MAXENT. For instance, consider the variable $X_j\in\bX$, and assume we are given the distribution $P(\bbX)$ for $\bbX=\bX\bsh\left\{X_j\right\}$. Additionally, we are given some expectations involving the variables $\bbX$ and $X_j$. In this case, we obtain the MAXENT solution for the joint distribution of $\bX$ by maximising the conditional entropy
\begin{align}
H_p(X_j \mid \bbX)&= -\sum_{\bx} p(x_j\mid\bbx)p(\bbx)\log p(x_j\mid\bbx) \; ,
\end{align} 
subject to the constraints in \cref{eq:maxent_constraints} imposed by the expectations of the functions $f$ as before, but now the set of variables $\bX_{S_k}$ the function $f_k$ acts upon always contains the variable $X_j$, so $\bX_{S_k}=\bbX_{S_k}\cup\left\{X_j\right\}$ for $\bbX_{S_k}\subseteq\bbX$. 
In this case, the solution in the Lagrangian formalism reads
\begin{align}
\hat{p}(x_j\mid\bbx) &=\exp\left[\sum_{k} \lambda_{k}f_{k}(\bx_{S_k}) -\beta(\bbx)\right] \; , \label{eq:cmaxent_sol_pyx}
\end{align}
where $\lambda$ are the respective Lagrange multipliers for which optimal values can be found analogously to \cref{eq:maxent_lambdas}, and $\beta(\bbx)=\log\sum_{x_j} \exp\left[\sum_{k} \lambda_{k}f_{k}(\bx_{S_k}) \right]$ ensures that the marginal constraint $\hat{p}(\bbx)=p(\bbx)$ is satisfied. The joint MAXENT distribution is then given by $\hat{p}(\bx)= \hat{p}(x_j\mid\bbx){p}(\bbx)$. 

{\bf Using conditional means} 
When we consider a scenario as described in the introduction, in which we want to merge the information from different studies or research papers, we might only be provided with {\it conditional means}, like the average depression rate given that the age is in a specific range. 
In this case, the given constraints would be 
\begin{align} \label{eq:cmaxent_constraints_conditional}
|\Exp{p}{f_k \mid \bbx_{S_k}=\bbx_{S_k}^\nu} - \tilde{f}_k^{\nu}|\leq\hat{\varepsilon}_k^\nu \quad \forall k,\nu \; ,
\end{align}
for $\nu=1,\dots,\mathcal{V}_k$ and $\bbx_{S_k}^1,\dots,\bbx_{S_k}^{\mathcal{V}_k}$ being the possible sets of values the set of discrete random variables $\bbX_{S_k}$ can attain. Then \cref{eq:cmaxent_constraints_conditional} replaces the constraints in \cref{eq:maxent_constraints} and the conditional MAXENT solution reads
\begin{align}\label{eq:cmaxent_sol_conditional}
\hat{p}(x_j\mid\bbx) = \exp\left[\sum_{k,\nu}\hat{\lambda}_k^{\nu} f_k(\bx_{S_k})\delta_{\bbx_{S_k},\bbx_{S_k}^\nu}-\hat{\beta}(\bbx) \right]  
\end{align}
with the Lagrange multipliers $\hat{\lambda}=\left\{\hat{\lambda}_k^\nu\right\}$ and 
$\hat{\beta}(\bbx)=\log\sum_{x_j} \exp\left[\sum_{k,\nu}\hat{\lambda}_k^{\nu}f_k(\bx_{S_k})\delta_{\bbx_{S_k},\bbx_{S_k}^\nu} \right]$ 
and 
\begin{align} \label{eq:delta}
\delta_{\bbx_{S_k},\bbx_{S_k}^\nu} = 
\begin{cases}
1 \quad &\text{if } \bbx_{S_k}=\bbx_{S_k}^\nu \\
0 \quad &\text{otherwise}  \; .
\end{cases}
\end{align}


\section{OBTAINING CAUSAL INFORMATION BY MERGING DATASETS WITH MAXENT} \label{sec:merging_datasets}
In this section, we consider the analysis of the causal relationship between variables if not all variables have been observed jointly. All proofs of the following propositions can be found in \cref{app:proofs}.

First, we show how to detect the presence or absence of direct causal links in a DAG $G$ from the Lagrange multipliers of the MAXENT distribution. 
We start by showing that if $X_i$ and $X_j$ are CI given all other variables w.r.t.\ the true distribution, then this is also the case w.r.t.\ the MAXENT distribution and reflects in the respective Lagrange multipliers being zero.

\begin{restatable}[CI results in Lagrange multipliers being zero]{Lemma}{cioneway}\label{th:ci_one_way}
Let $P$ be a distribution and let $\hat{P}$ be the MAXENT distribution satisfying the constraints imposed by the expectations of the functions $f$ which are sufficient to uniquely describe the marginal distributions $P(X_i,\bZ), P(X_j,\bZ)$, and $P(X_i,X_j)$. Then it holds:
\begin{align}\label{eq:ci_one_way}
&X_i\CI X_j \mid \bZ \;\; [P] \notag\\ 
\quad\Rightarrow\quad &X_i\CI X_j\mid \bZ \;\; [\hat{P}] \notag\\
 \quad \Rightarrow \quad &\lambda_k = 0 \quad \forall k\;\text{ with }\; \bX_{S_k}=\left\{X_i,X_j\right\} \; .
\end{align}
\end{restatable}

Under the stated assumptions, it directly follows from \cref{th:ci_one_way} that if two variables are CI given all other variables, and hence not directly linked in the causal DAG $G$, then the respective Lagrange multipliers are zero. This, however, is not enough to draw conclusions from the Lagrange multipliers about the absence or presence of causal links. 
For this, we first need to show that the presence of a direct link results in a non-zero Lagrange multiplier. But to do this, we first need to postulate a property that we call {\it faithful $f$-expectations}. 
This property is analogous to faithfulness in postulating the genericity of parameters. 
For the following definition, we denote with $\lambda^P_f$ and $\lambda_f^Q$ the set of Lagrange multipliers of the MAXENT distribution satisfying the expectation constraints in \cref{eq:maxent_constraints} entailed by the functions $f$ w.r.t.\ the distributions $P$ and $Q$, respectively. 

\begin{Definition}[Faithful $f$-Expectations]
A distribution $P$ is said to have faithful $f$-expectations relative to a DAG $G$, if $\lambda_{f_k}^P\neq 0$ for all $f_k\in f$ where it exists a distribution $Q$ that is Markov relative to $G$ and for which it is $\lambda_{f_k}^Q\neq 0$.
\end{Definition}

\begin{figure}[t]
\centering
\includegraphics[width=\columnwidth]{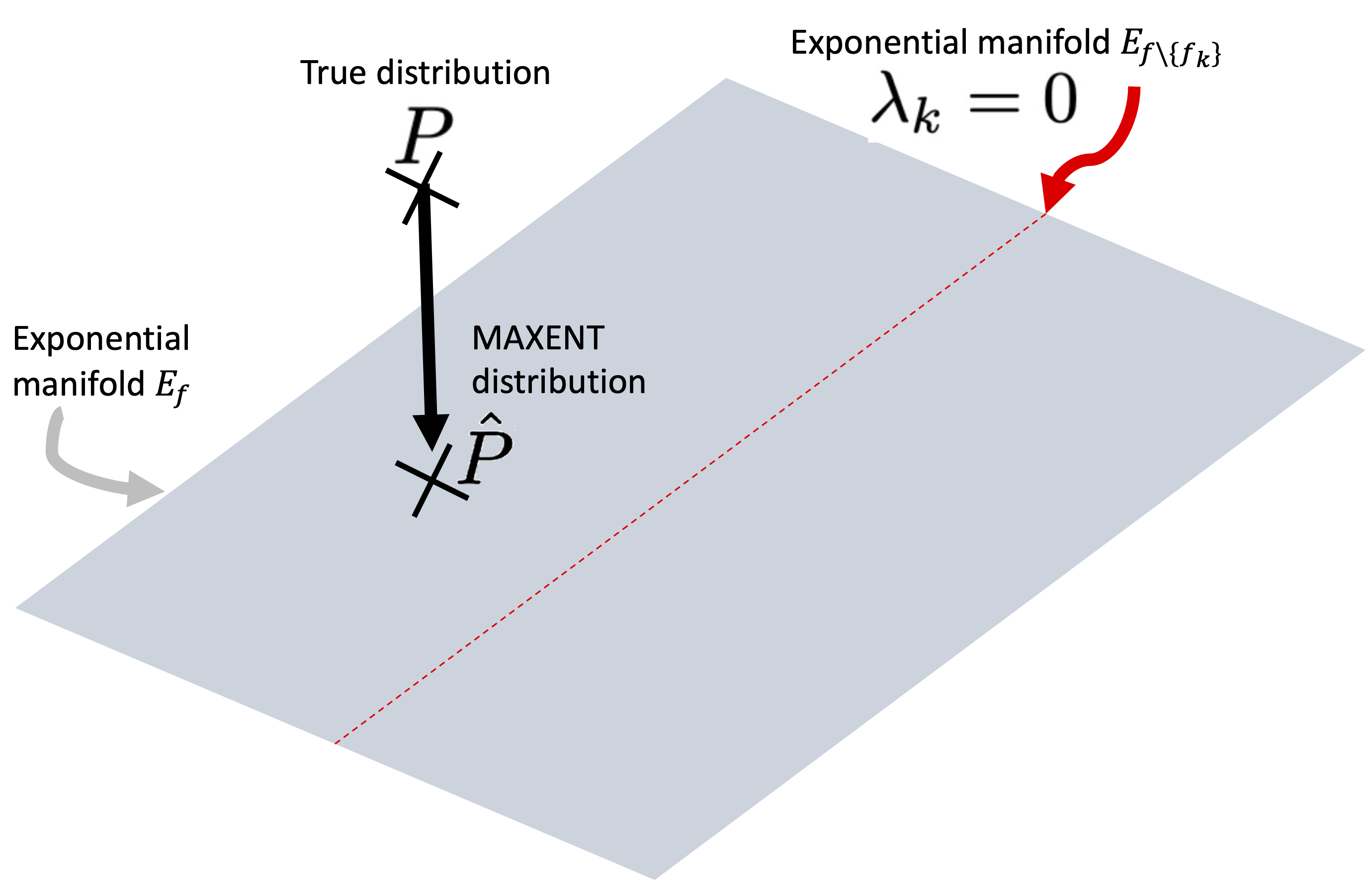}
\caption{Intuitive explanation of the idea behind faithful $f$-expectations: the MAXENT distribution is a projection of the distribution $P$ onto the exponential manifold $E_f$, defined by the span of the functions $f$. It is very unlikely that this projection falls into the submanifold $E_{f\setminus \left\{f_k\right\}}$ where $\lambda_k=0$ just by chance. \label{fig:faithfulness}}
\end{figure}

We rephrase this definition in the language of information geometry to show that this is just a genericity assumption like usual faithfulness, and \cref{fig:faithfulness} illustrates the intuition behind it. By elementary results of information geometry \citep{Amari}, the MAXENT distribution $\hat{P}$ can also be considered a projection of the distribution $P$ onto the exponential manifold $E_f$, which is defined by the span of all functions $f$, containing distributions of the form $\exp\left[\sum_k\lambda_k f_k(\bx_{S_k}) - \alpha\right]$ (visualised by the blue plane in \cref{fig:faithfulness}). If a Lagrange multiplier $\lambda_k$ is zero, then $\hat{P}$ lies within the submanifold $E_{f\setminus \{f_k\}} \subset E_f$ which is defined through the span of all functions $f$ without $f_k$ (illustrated by the red, dashed line in \cref{fig:faithfulness}). Then faithful $f$-expectations state that the projection of $P$ onto $E_f$ will generically not lie in $E_{f\setminus  \{f_k\}}$ unless the DAG $G$ only allows for distributions whose projections onto $E_f$ also lie in $E_{f\setminus  \{f_k\}}$. 

Further justification of faithful $f$-expectations via some probabilistic arguments would be a research project in its own right. After all, even the discussion on usual faithfulness is ongoing:  The \enquote{measure zero argument} by \citet{Meek} is criticised in \citet{LemeireJ2012}, and it is argued that natural conditional distributions tend to be more structured. In \citet{uhler2013} it is shown that distributions are not unlikely to be {\it close to} being unfaithful. Despite these concerns, faithfulness still proved to be helpful.

Postulating faithful $f$-expectations allows us to link the causal structure to the Lagrange multipliers:

\begin{restatable}[Causally linked variables have non-zero Lagrange multipliers]{Lemma}{fexpectations}\label{lm:f-expectations}
Let $P$ have faithful $f$-expectations relative to a causal DAG $G$. Then it is $\lambda_k^P\neq 0$ for any bivariate function $f_k$ whose variables are connected in $G$.
\end{restatable}

Now we have all we need to connect the structure of the causal DAG and the Lagrange multipliers. 

\begin{restatable}[Causal structure from Lagrange multipliers]{Theorem}{cibothways}\label{th:ci_both_ways}
Let $P$ be a distribution with faithful $f$-expectations w.r.t.\ a causal DAG $G$, and let $\hat{P}$ be the MAXENT solution satisfying the constraints imposed by the expectations of the functions $f$ which are sufficient to uniquely describe the marginal distributions $P(X_i,\bZ),P(X_j,\bZ)$, and $P(X_i,X_j)$. Then the following two statements hold: 
\begin{enumerate}
\item If $\bZ$ is d-separating $X_i$ and $X_j$ in $G$, then all Lagrange multipliers $\lambda_k$ are zero for all $k$ with $\bX_{S_k}=\left\{X_i,X_j\right\}$. 
\item If $\lambda_k=0$ for all $k$ with $\bX_{S_k}=\left\{X_i,X_j\right\}$, then there is no direct link between $X_i$ and $X_j$ in the DAG $G$.
\end{enumerate}
\end{restatable}

For the special case where we have some prior knowledge about the causal order, e.g.\ if we know that $X_j$ can be causally influenced by $X_i$ or $\bZ$, but not the other way around, we can directly identify the absence or presence of a direct causal link between $X_i$ and $X_j$: 

\begin{restatable}[Identification of causal links when causal order is known]{Corollary}{noedge}\label{lm:no_edge}
Let $P$ be a distribution with faithful $f$-expectations w.r.t.\ a causal DAG $G$, and let $\hat{P}$ be the MAXENT solution satisfying the constraints imposed by the expectations of the functions $f$ which are sufficient to uniquely describe the marginal distributions $P(X_i,\bZ),P(X_j,\bZ)$, and $P(X_i,X_j)$. If it is excluded that $X_j$ can causally influence $X_i$ and $\bZ$, i.e.\ the DAG $G$ cannot contain edges $X_j\to X_i$ or $X_j\to\bZ$, then it holds
\begin{align} 
&X_i \text{ is not directly linked to } X_j \notag\\
\quad \Leftrightarrow \quad &\lambda_k = 0 \quad \forall k\; \text{ with }\; \bX_{S_k}=\left\{X_i,X_j\right\}\; .
\end{align}
This also holds for conditional MAXENT, and if conditional means are used (see \cref{eq:cmaxent_constraints_conditional}) it holds
\begin{align} 
&X_i \text{ is not directly linked to } X_j \notag\\
\Leftrightarrow \quad &\hat{\lambda}_k^\nu=\hat{\lambda}_k^{\nu'}  \quad\forall\nu,\nu', k\text{ with } \bX_{S_k}=\left\{X_i,X_j\right\}\; .
\end{align}
\end{restatable}

Note that when conditional means are used, 
the Lagrange multipliers need not be zero to indicate missing links, but need to be constant for all conditions. 
We use this result in our experiments in \cref{sec:experiments}, 
where we estimate conditional MAXENT in the causal order, 
which is called \textit{causal MAXENT}, as proposed and justified by \cite{janzingPIR}.

The reader may wonder about more general statements like the question \enquote{What information can be obtained about a DAG with $N$ nodes if only bivariate distributions are available?}. 
For this scenario, we have at least a necessary condition for causal links. For this recall that for any DAG $G$, the corresponding {\it moral graph} $G^m$ is defined as the undirected graph having edges if and only if the nodes are directly connected in $G$ or have a common child \citep{Lauritzen}. 

\begin{restatable}[Graph constructed from MAXENT with only bivariate constraints is a supergraph of the moral graph]{Theorem}{moral}\label{thm:moral}
Let $f$ be a basis for the space of univariate and bivariate functions, i.e.\ the set of $f$-expectations determine all bivariate distributions
uniquely. Let $P$ be a joint distribution that has faithful $f$-expectations w.r.t.\ the DAG $G$. 
Let $G^b$ be the undirected graph constructed from the MAXENT distribution by connecting $X_i$ and $X_j$ 
if and only if  there is a non-zero Lagrange multiplier corresponding 
to some bivariate function of $X_i$ and $X_j$. Then $G^b$ is a supergraph of $G^m$, the moral graph of $G$. 
\end{restatable} 

\Cref{thm:moral} provides at least a candidate list for potential edges from bivariate information alone, which tells us where additional observations are needed to identify edges. The edges are candidates for being in the Markov blanket, which thus limits the number of variables that need to be considered for a prediction model.

Note that inferring causal relations via bivariate information is not uncommon: after all, many implementations of the PC algorithm \citep{Spirtes2000,kalisch2007estimating,kalisch2008robustification,harris2013pc,cui2016copula,tsagris2019bayesian} use partial correlations instead of CIs. For real-valued variables, one can interpret this in the spirit of this paper since it infers CIs to hold whenever they are true for the multivariate Gaussian
matching the observed first and second moments (i.e.\ the unique MAXENT distribution satisfying these constraints). 
These heuristics avoid the complex problem \citep{Shah2020} of non-parametric CI testing. In addition to the results above, our approach also generalises the partial correlation heuristics to more general functions $f_k$, including multivariate and higher-order statistics.   

Note also that we do not propose a {\it general purpose} conditional independence test because we do not fully understand what sort of conditional dependence it detects. We have concluded that it \enquote{generically} (i.e.\ subject to faithful $f$-expectations) has power against conditional dependencies generated by a DAG. Without a DAG (whose distributions can be easily parameterised), we do not see a clear notion of genericity on which a similar statement could be based.

For most applications, estimating the joint distribution of many variables is not an end in itself. Instead, one will often be interested in particular properties of
the joint distributions for specific reasons. In these cases, MAXENT is already helpful if it resembles the statistical properties of interest. So far, we have
shown this for some conditional independence. We will now sketch how entropy maximisation can be used for pooling predictions made from different datasets.

\begin{restatable}[Predictive power of MAXENT]{Theorem}{predictive}\label{th:predictive_power}
Let $X_j, X_i, \bZ$ be binary variables, with $\bZ$ possibly high dimensional.
Furthermore, let $\hat{P}(X_j \mid X_i, \bZ)$ be the MAXENT solution that maximises the conditional entropy of $X_j$ given $X_i$ and $\bZ$,
subject to the moment constraints given by the observed pairwise distributions $P(X_j, X_i)$, $P(X_j, \bZ)$, and $P(X_i, \bZ)$.
Then $\hat{P}$ is a better predictor of $X_j$ than any of the individual bivariate probabilities, as measured by the likelihood of any point where all variables are observed, i.e.\ a point from $P(X_j, X_i, \bZ)$.
\end{restatable}

This show that merging datasets with MAXENT also improves the predictive power compared to using the observed marginal distributions.


\section{RELATED WORK}\label{sec:related_work}

In the context of missing data \citep{Rubin1976,Bareinboim2011,Mohan2021}, many methods have been developed to investigate CI, how the joint distribution of the variables factorises, and how to predict the result of interventions \citep{Pearl2000,Spirtes2000,Chickering2002b,tsamardinos2006max}. 
However, most of these approaches assume they are given one dataset in which some values are missing at random. 
More recently, the even more challenging task of inferring causal relationships from multiple datasets \citep{tillman2009structure,ramsey2010six,triantafillou2010learning,eberhardt2010combining,claassen2010causal,tillman2011learning,hyttinen2013discovering,tillman2014learning}. 
These approaches and our method have in common that they assume that the underlying causal structures are similar across the different datasets. 
\citet{triantafillou2010learning,tillman2011learning}, for instance, assume that a single causal mechanism generates the data and that the dependencies and independencies are captured by a maximal ancestral graph (MAG) and the m-separation criterion \citep{richardson2002ancestral}. 
Various methods have also been proposed to discover the causal graph from multiple datasets containing measurements in different environments. 
Some combine statistics or constraints from the different datasets to construct a single causal graph \citep{claassen2010causal,tillman2011learning,hyttinen2013discovering,hyttinen2014constraint,triantafillou2015constraint,rothenhausler2015backshift,forre2018constraint}, while others directly combine the data from the different datasets and construct a causal graph from the pooled data \citep{cooper1997simple,hauser2012characterization,cooper2013causal,mooij2013cyclic,peters2016causal,oates2016estimating,zhang2017causal,mooij2020joint}. 
In \citet{mooij2020joint}, for instance, the union of causal graphs in each dataset (or context) is found by jointly modelling the context variables and the observed variables. 
The main difference between these approaches and ours is that they all rely on statistical information that reveals CIs in each dataset individually. 
Hence they can only be applied if at least three variables have been observed jointly, while our approach can also be used if only pairwise observations are available.

In \cite{gresele2022causal}, the structural marginal question was asked: Can marginal causal models over subsets of variables with known causal graph be consistently merged? They proved that certain SCM can be falsified using only interventional and observetional data and a known graph structure. Their work differs from ours in that we are interested in interventional quantities, while they focus on counterfactual ones. As a result, the questions that can be answered with our framework are different.

In \cref{app:add_related_work} we comment on less related -- but still interesting -- literature on gaining statistical information from causal knowledge and other entropy-based approaches to extract and exploit causal information.


\section{EXPERIMENTS}\label{sec:experiments}

\begin{figure*}[!ht]
\sbox0{\begin{subfigure}[t]{.3\textwidth}
    \centering
    \adjustbox{max width=.9\textwidth}{

\begin{tikzpicture}
\node[latent] (U) {$U_1$};
\node[obs,below=1cm of U] (X1) {$X_1$};
\node[obs,right=.5cm of X1] (X2) {$X_2$};
\node[obs,right=.5cm of X2] (X3) {$X_3$};
\node[obs,right=.5cm of X3] (X4) {$X_4$};
\node[obs,right=.5cm of X4] (X5) {$X_5$};
\node[latent,above=1cm of X4] (V) {$U_2$};
\node[obs,below=1cm of X3] (X0) {$X_0$};

\edge{U}{X1,X2};
\edge{V}{X3,X4};
\edge[dashed]{X1}{X0};
\edge[dashed]{X2}{X0};
\edge[dashed]{X3}{X0};
\edge[dashed]{X4}{X0};
\edge[dashed]{X5}{X0};

\end{tikzpicture}}
    \subcaption{Structure graph (a)\label{fig:graph_exp_a}}
\end{subfigure}}
\sbox1{\begin{subfigure}[t]{.3\textwidth}
    \centering
    \adjustbox{max width=.9\textwidth}{

\begin{tikzpicture}
\node[latent] (U1) {$U_1$};
\node[latent,right=.5cm of U1] (U2) {$U_2$};
\node[latent,right=.5cm of U2] (U3) {$U_3$};
\node[latent,right=.5cm of U3] (U4) {$U_4$};
\node[latent,right=.5cm of U4] (U5) {$U_5$};
\node[obs,below=1cm of U1] (X1) {$X_1$};
\node[obs,below=1cm of U2] (X2) {$X_2$};
\node[obs,below=1cm of U3] (X3) {$X_3$};
\node[obs,below=1cm of U4] (X4) {$X_4$};
\node[obs,below=1cm of U5] (X5) {$X_5$};
\node[obs,below=1cm of X3] (X0) {$X_0$};

\edge{U1}{X1, X2, X3};
\edge{U2}{X2, X3, X4};
\edge{U3}{X3, X4, X5};
\edge{U4}{X4, X5, X1};
\edge{U5}{X5, X1, X2};
\edge[dashed]{X1}{X0};
\edge[dashed]{X2}{X0};
\edge[dashed]{X3}{X0};
\edge[dashed]{X4}{X0};
\edge[dashed]{X5}{X0};

\end{tikzpicture}}
    \subcaption{Structure graph (b)\label{fig:graph_exp_b}}
\end{subfigure}}
\sbox2{\begin{subfigure}[t]{.3\textwidth}
    \centering
    \adjustbox{max width=.9\textwidth}{

\begin{tikzpicture}
\node[latent] (U) {$U_1$};
\node[obs,below=1cm of U] (X3) {$X_3$};
\node[obs,right=.5cm of X3] (X5) {$X_5$};
\node[obs,left=.5cm of X3] (X1) {$X_1$};
\node[obs,left=.5cm of X1] (X2) {$X_2$};
\node[obs,right=.5cm of X5] (X4) {$X_4$};
\node[obs,below=1cm of X3] (X0) {$X_0$};

\edge{U}{X1, X2, X3, X4, X5};
\edge{X1}{X2, X3};
\edge{X5}{X3, X4};
\edge[dashed]{X1}{X0};
\edge[dashed]{X2}{X0};
\edge[dashed]{X3}{X0};
\edge[dashed]{X4}{X0};
\edge[dashed]{X5}{X0};

\end{tikzpicture}}
    \subcaption{Structure graph (c)\label{fig:graph_exp_c}}
\end{subfigure}}
\sbox3{\begin{subfigure}[t]{.25\textwidth}
        \centering
        \adjustbox{max width=\textwidth}{\includegraphics[width=\textwidth]{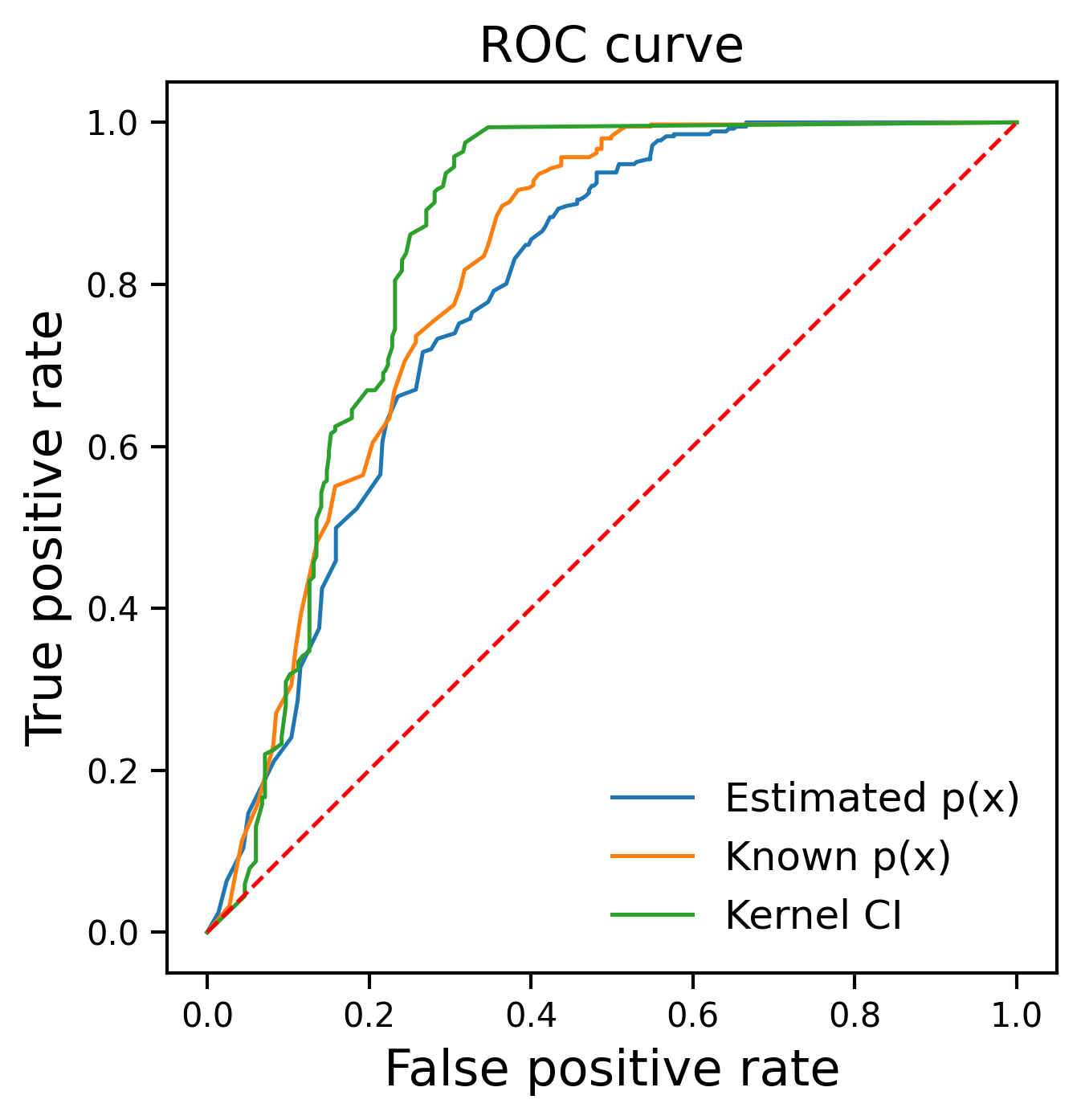}}
        \caption{ROC curve for graph (a) \label{fig:roc_overlay_a}}
\end{subfigure}}
\sbox4{\begin{subfigure}[t]{.25\textwidth}
	    \centering
    \includegraphics[width=\textwidth]{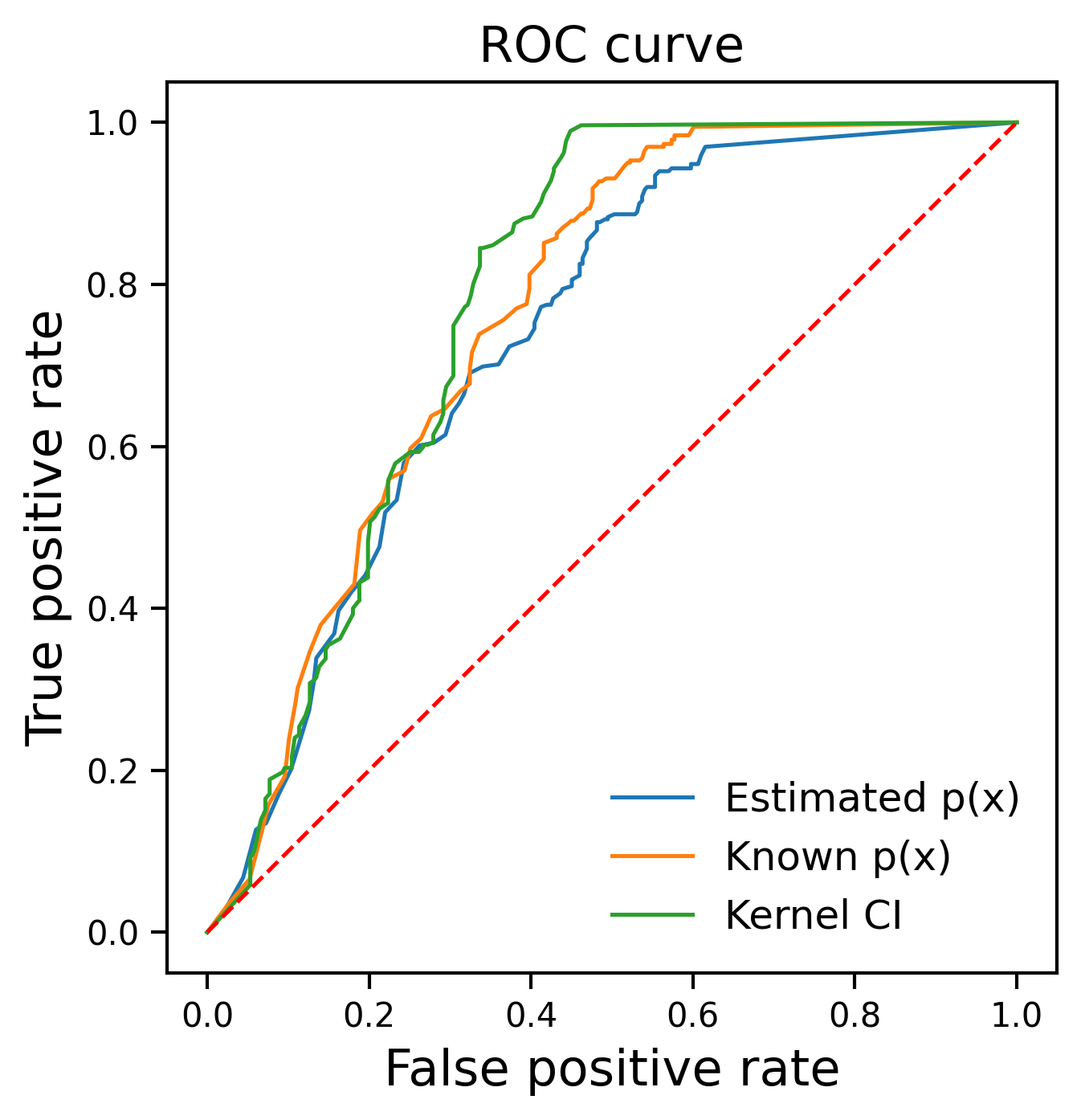}
    \caption{ROC curve for graph (b) \label{fig:roc_overlay_b}}
\end{subfigure}}
\sbox5{\begin{subfigure}[t]{.25\textwidth}
	    \centering
    \includegraphics[width=\textwidth]{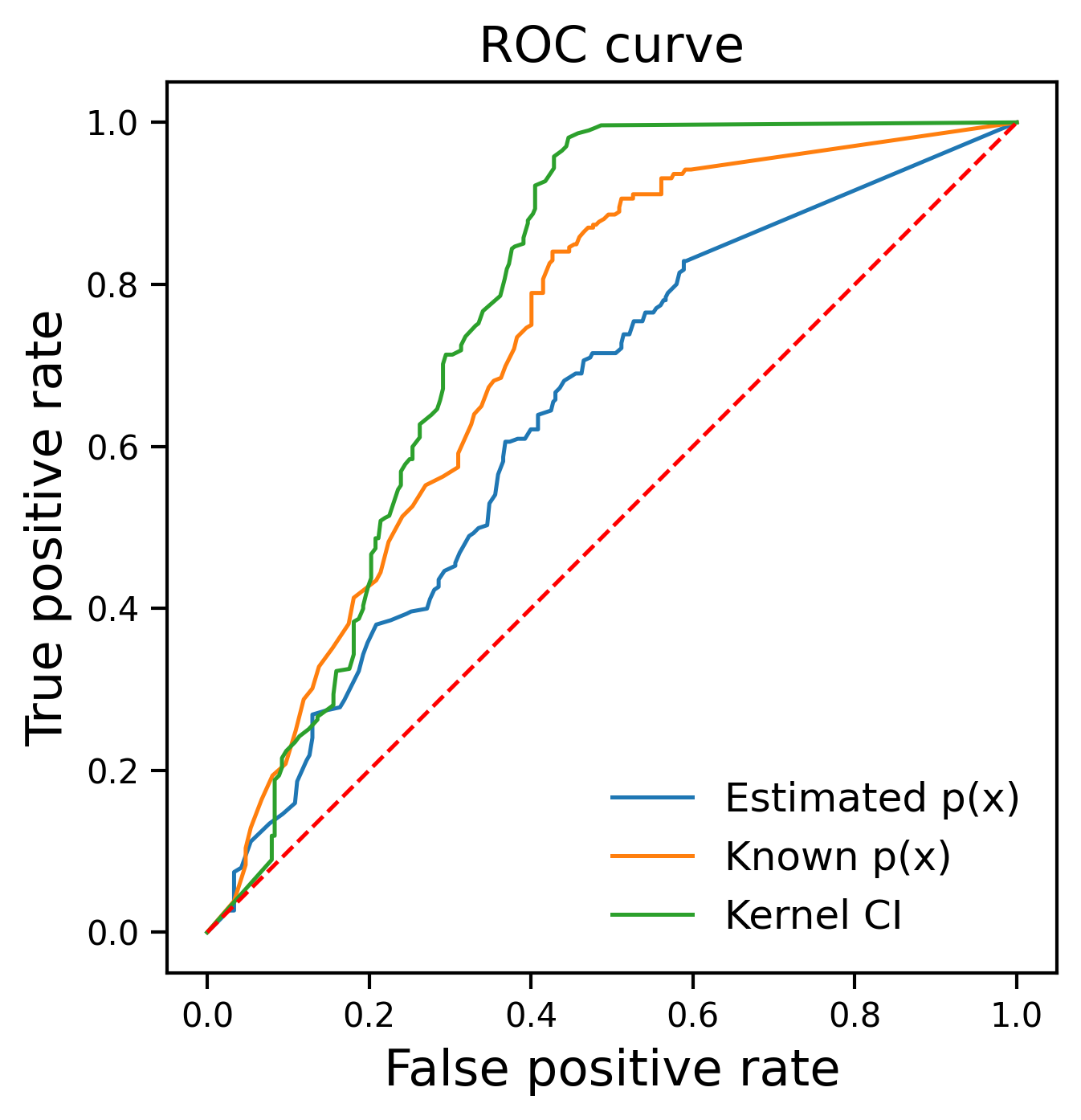}
    \caption{ROC curve for graph (c) \label{fig:roc_overlay_c}}
\end{subfigure}}
\sbox6{\begin{subfigure}[t]{.32\textwidth}
        \centering
        \adjustbox{max width=\textwidth}{\includegraphics[width=\textwidth]{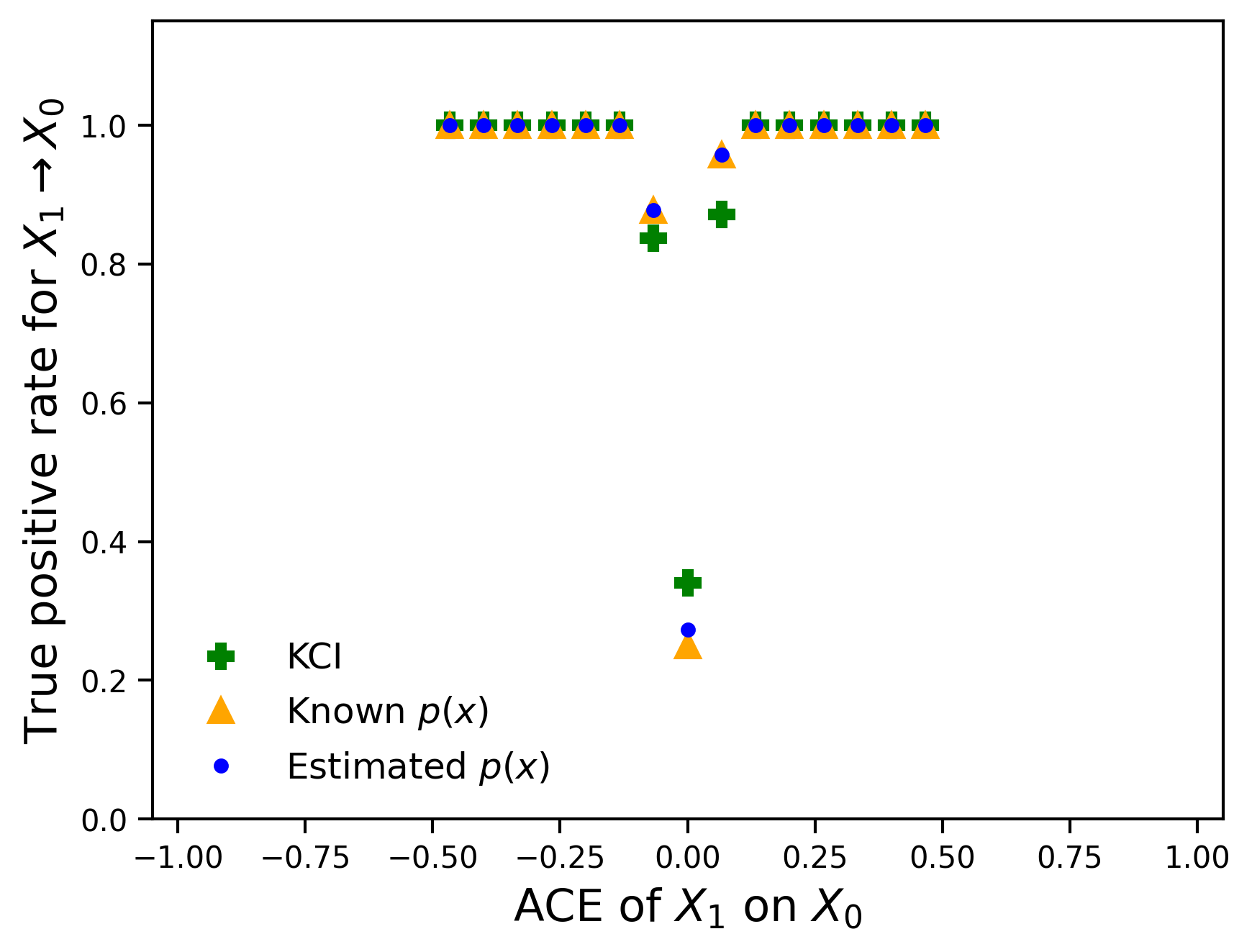}}
        \caption{True positives over ACE, graph (a) \label{fig:rate_vs_strength_a}}
\end{subfigure}}
\sbox7{\begin{subfigure}[t]{.32\textwidth}
	    \centering
    \includegraphics[width=\textwidth]{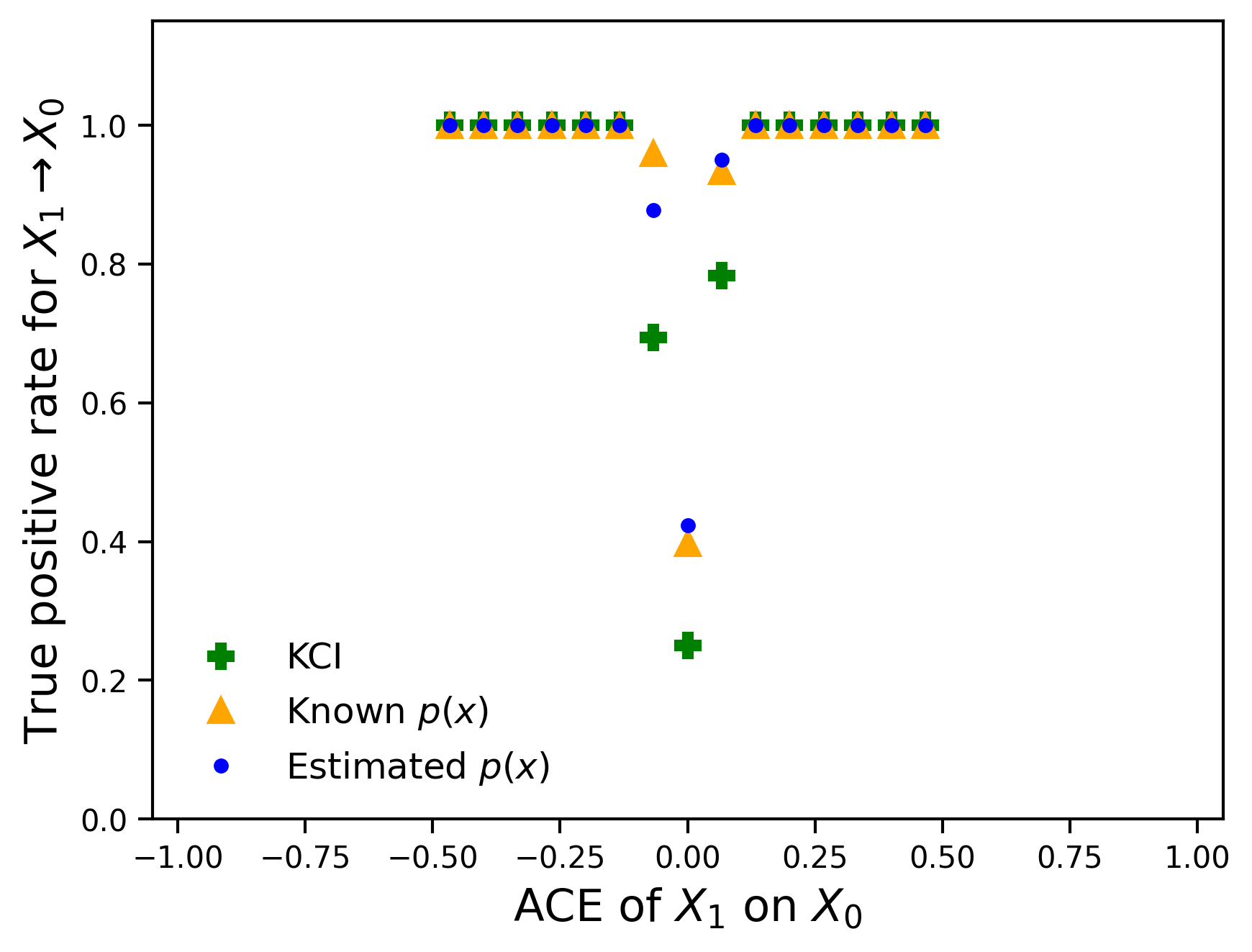}
    \caption{True positives over ACE, graph (b) \label{fig:rate_vs_strength_b}}
\end{subfigure}}
\sbox8{\begin{subfigure}[t]{.32\textwidth}
	    \centering
    \includegraphics[width=\textwidth]{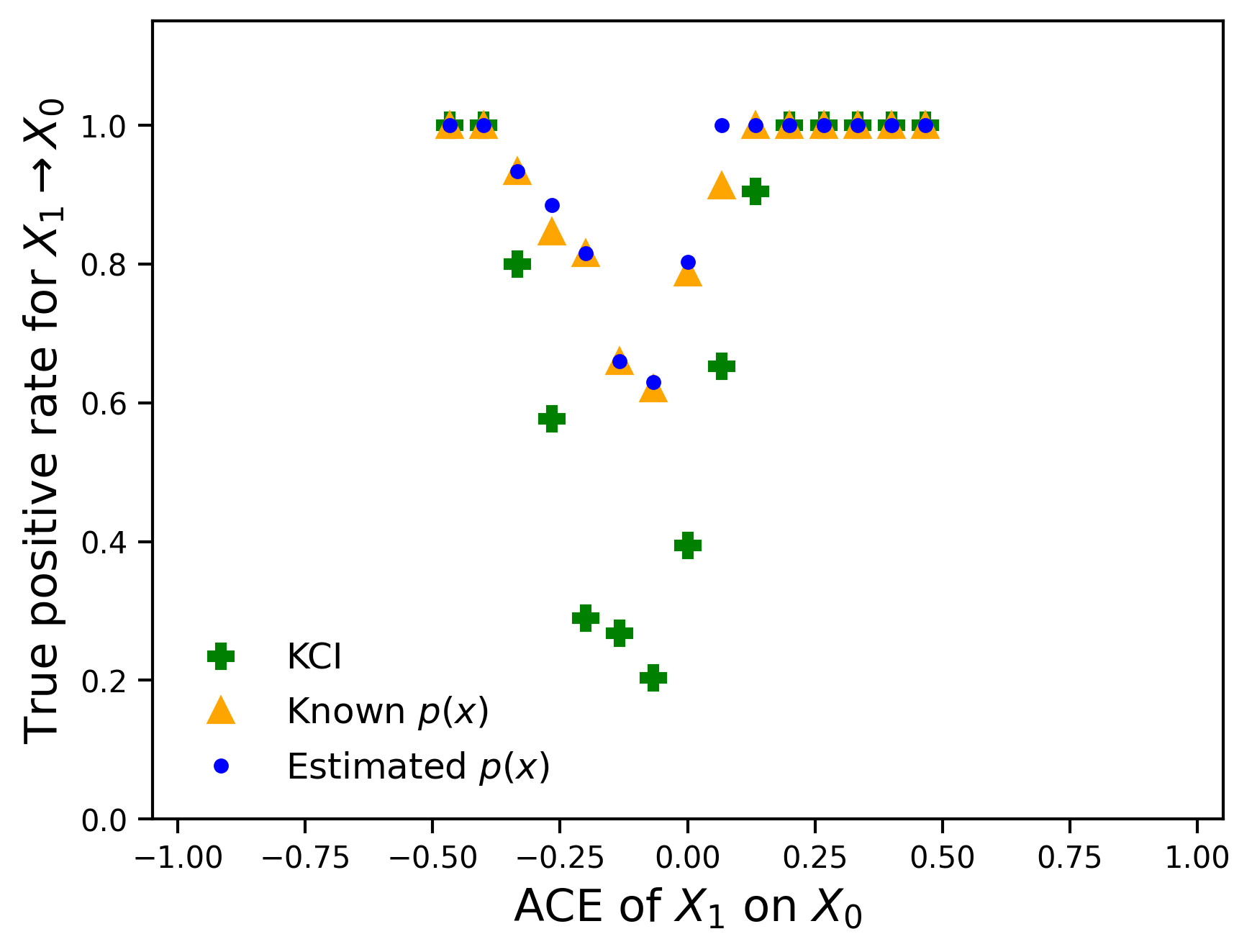}
    \caption{True positives over ACE, graph (c) \label{fig:rate_vs_strength_c}}
\end{subfigure}}
\centering
\begin{tabular}{ccc}
\usebox0  & \usebox1 & \usebox2 \\[2em]
\usebox3  & \usebox4 & \usebox5 \\[2em]
\usebox6  & \usebox7 & \usebox8 \\[2em]
\end{tabular}
\caption{
    We show the structure of the graphs we consider in the synthetic experiments in (\subref{fig:graph_exp_a}), (\subref{fig:graph_exp_b}), and (\subref{fig:graph_exp_c}). 
    In (\subref{fig:roc_overlay_a}), (\subref{fig:roc_overlay_b}), and (\subref{fig:roc_overlay_c}) we show the ROC curves for the identification of missing edges. We generated 100 datasets for each graph where we varied the used SCMs and the absence and presence of the edges shown as dashed lines, whereas links represented by solid lines are always present.
    In (\subref{fig:rate_vs_strength_a}), (\subref{fig:rate_vs_strength_b}), and (\subref{fig:rate_vs_strength_c}) we show how the ability to detect an edge depends on the strength of the causal effect. Here we generated another 500 datasets for each graph in which the link between $X_1$ and $X_0$ was always present, but the ACE of $X_1$ on $X_0$ varied. 
    Although our MAXENT-based approach only uses conditional means as input, it achieves similar performance as the KCI-test that uses the full generated dataset.
    \label{fig:exp_results}}
\end{figure*}

In this section, we apply the theoretical results from \cref{sec:merging_datasets} on different synthetically generated and real-world datasets. 
For this, we implemented the MAXENT estimation in Python (see \cref{sec:implementation}).

{\bf Synthetic data}
We consider five binary variables $X_1,\dots,X_5$, which are potential causes of a sixth binary variable $X_0$, and we want to infer which variables $X_i$ have a direct causal link to $X_0$. 
The ground truth DAGs for our experiments are shown in \cref{fig:graph_exp_a,fig:graph_exp_b,fig:graph_exp_c}, and the SCMs we used for the data generation can be found in \cref{sec:experimental_setup}. 

For the first set of experiments, we kept the structure of the confounders $U_j$ with the potential causes fixed (solid lines) and randomised the existence of mechanisms between the potential parents and the effect variable $X_0$ (dashed lines).
We generated 100 datasets for each graph structure by randomly picking the existing mechanisms and the parameters used in the SCM.
We sample 1000 data points according to the respective SCM for each dataset. Then we artificially split these observations into five datasets that we want to merge and that always only contain bivariate information about $X_0$ and one of the potential causes $X_i$. We do this by empirically estimating the conditional means $\Exp{p}{x_0\mid x_i=0}$ and $\Exp{p}{x_0\mid x_i=1}$ from the samples for all $i=1,\dots,5$. We use these conditional means as constraints for the MAXENT optimisation problem as shown in \cref{eq:cmaxent_constraints_conditional}. 
We assume that $X_0$ cannot have a causal influence on any $X_i$. Therefore, we can use the results in \cref{lm:no_edge} to identify whether $X_i$ is directly causally linked to $X_0$ or not. 
To decide whether the Lagrange multipliers associated with a potential cause $X_i$ are constant -- and hence $X_i$ is not directly linked to $X_0$ -- we use a relative difference estimator 
\begin{align}
\theta_i = \frac{\left|\lambda_i^{1}-\lambda_i^{2}\right|}{\max\{|\lambda_i^{1}|, |\lambda_i^{2}|, \left|\lambda_i^{1}-\lambda_i^{2}\right|, 1\}} \;\in\left[0,1\right] \; ,
\end{align}
where $\lambda_i^1, \lambda_i^2$ are the two Lagrange multipliers for the constraints associated with $X_i$. 
We consider the Lagrange multipliers constant if $\theta_i$ is smaller than a threshold $t \in [0, 1]$. 
We vary the threshold $t$ linearly between zero and one. 
We count the number of correctly and falsely identified edges in the 100 datasets for each threshold value.
The results are summarised in the receiver operating characteristic (ROC) curves in \cref{fig:roc_overlay_a,fig:roc_overlay_b,fig:roc_overlay_c}. 
We consider two scenarios: one in which we assume that we know the marginal distribution $P(X_1,\dots,X_5)$ for the potential causes (called \enquote{known $p(x)$}, orange line), and the second where we first infer this distribution also using MAXENT (called \enquote{estimated $p(x)$}, blue line). Further, we compare our results with a kernel-based conditional independence test (KCI-test) \citep{UAI_Kun_kernel,strobl2019approximate} (green line). For the KCI-test, we directly use the 1000 data points generated from the joint distribution. To generate the ROC curve, we vary the $\alpha$-level of the test for the null hypothesis that $X_0$ is CI of $X_i$ given all other potential causes and count the number of correct/false rejections/acceptances.

In the second set of experiments, we investigate how much our approach's ability to identify edges depends on the strength of the causal effect. We generated 500 additional datasets for each graph as described before, but this time always included a causal link from $X_1$ to $X_0$ and only varied the strength of this connection. We fixed the threshold for the identification of an edge to a randomly picked value ($t=\alpha=0.15$). \Cref{fig:rate_vs_strength_a,fig:rate_vs_strength_b,fig:rate_vs_strength_c} show how in this case the true positive rate for the identification of the link depends on the ACE of $X_1$ on $X_0$. 

The results in \cref{fig:roc_overlay_a,fig:roc_overlay_b,fig:roc_overlay_c,fig:rate_vs_strength_a,fig:rate_vs_strength_b,fig:rate_vs_strength_c} show that for all graph structures our method achieves similar performance as the KCI-test. This is impressive, as our method only uses the conditional means of $X_0$ on only one of the potential causes. In contrast, the KCI-test uses all samples generated from the joint data distribution. That means that, although our method uses much less information than the KCI-test and even merges these little pieces of information from different datasets, our method still achieves similar performance as the KCI-test. 

\begin{figure}[t]
\centering
\includegraphics[width=.9\columnwidth]{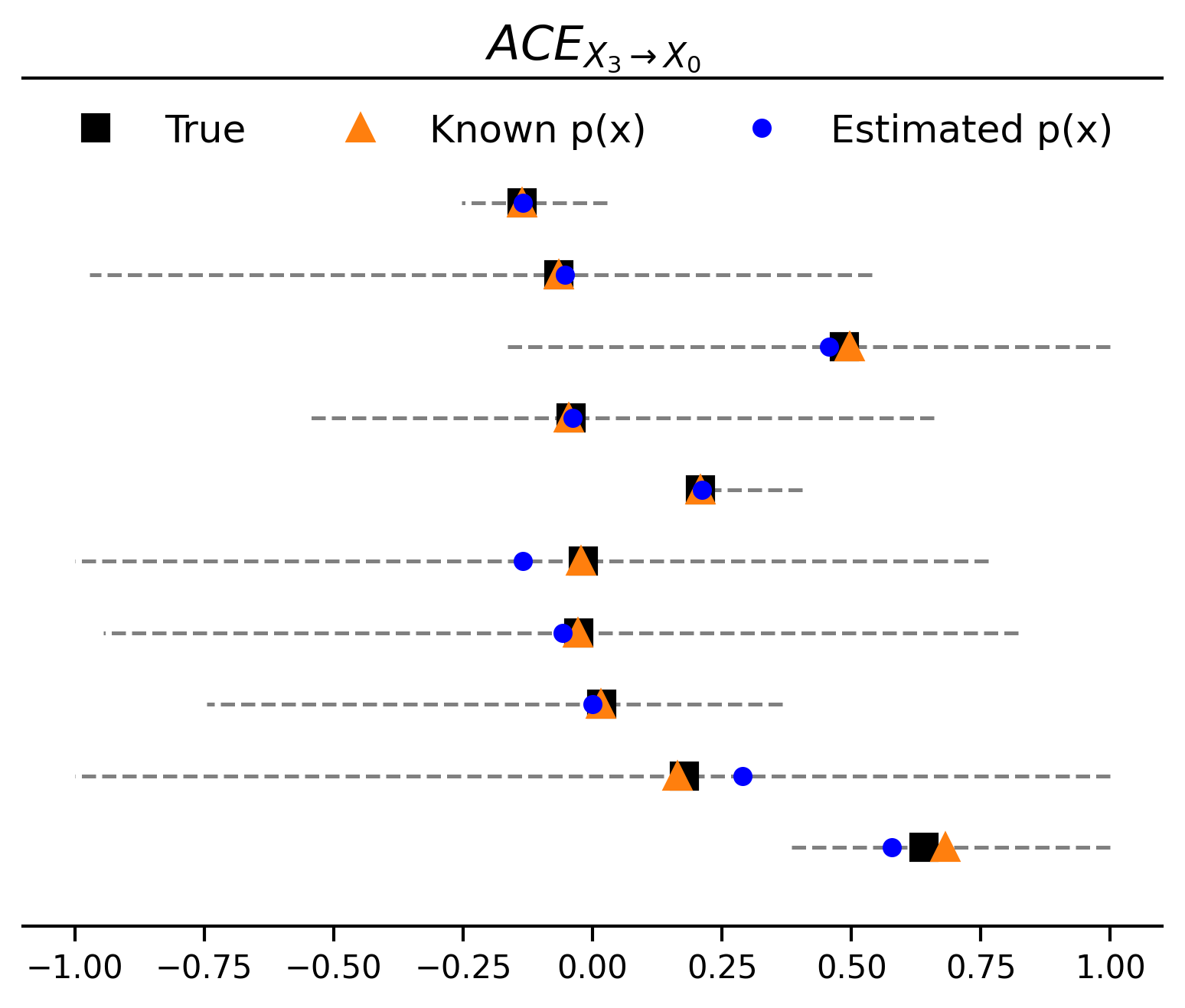}
\caption{ACE of $X_3$ on $X_0$ for ten randomly picked examples of a variation of the graph in \cref{fig:graph_exp_c}. The ACE estimated from the MAXENT solution with a known marginal distribution of the causes (orange triangle) is always close to the true ACE (black square). But even when the distribution of the causes is also inferred using MAXENT (blue dot), the estimated ACE is close to the true one and always within the bounds (grey lines) estimated from the marginal distributions. \label{fig:ace}}
\end{figure}

In addition, we want to show that the MAXENT solution can not only provide information about the causal structure but even about the strength of a causal effect. For this, we derive bounds for the ACE based only on the marginal distributions in \cref{sec:causal_influence}. In \cref{fig:ace} we see that the ACE estimated based on the MAXENT distribution is always very close to the true ACE, and even in the cases where they do not precisely coincide, they are both clearly within the bounds derived based on the marginal distributions.  

{\bf Real data}
We performed an experiment using real-world data from \citet{Gapminder}, a website that compiles country-level data of social, economic, and environmental nature. We chose three variables for our experiment: 
CO2 tonnes emission per capita \citep{osti_1389331}; 
inflation-adjusted Gross Domestic Product (GDP) per capita \citep{WB:GDPPC}; and
Human Development Index (HDI) \citep{UNHDR:HDI}. 
We use data from 2017 for all variables and standardise it before estimation. We consider CO2 emissions the target variable for which the other variables are potential causes. We use the unconditional mean and variance of CO2 emissions and the pairwise covariance between CO2 emissions and each of the two other variables as constraints. 

According to the Lagrange multipliers shown in \cref{tab:pvalues}, and \cref{lm:no_edge}, we conclude that CO2 is directly linked to HDI but not to GDP. 
We run the same KCI-test as in the synthetic experiments above to investigate this conclusion. \Cref{tab:pvalues} also shows the obtained p-values for the null hypothesis that CO2 emission is CI of each variable conditioned on the other variable. The results of the KCI-test agree with our conclusion. 
The result of the KCI-test, nonetheless, does not necessarily reflect the ground truth. However, GDP only has an indirect causal influence on CO2 emissions through HDI matches our intuition. We would expect that a change in GDP not directly affects the CO2 emissions but influences the HDI -- and potentially multiple other factors that we do not consider in this experiment -- that then affects the CO2 emissions. In \cref{sec:add_results} we discuss more such experiments in which we also include fertility and life expectancy as potential causes. In all of the considered cases where we used two potential causes, the conclusion drawn from the Lagrange multipliers agreed with the conclusion drawn from the KCI-test. Only when we include more variables, the KCI-test indicates CI of HDI and CO2 emissions given the other variables, while our method still finds a direct link between them. However, this finding of the KCI-test is also inconsistent with the other CI statements of the KCI-test for smaller conditioning sets. 
Moreover, consider what is available to both methods: 
The KCI-test requires a sample of the joint distribution, while our method relies solely on bivariate covariances, which might not even be enough to describe the joint distribution fully. 

\begin{table}[t]
    \centering
    \caption{Found Lagrange multipliers $\lambda_i$ for the MAXENT solution and the p-values for the KCI-test. We indicate where the multipliers / p-values indicate the presence of a direct edge connecting $X_i$ and CO2 emissions / that the two are not CI given the other variable. \label{tab:pvalues}}
    \begin{adjustbox}{max width=\columnwidth}
    \begin{tabular}{lrc|rc}
    \toprule
    variable $X_i$ & $\lambda_i$ & edge & p-value & no CI \\
    \midrule
     GDP & -0.29 & \ding{55} & 0.19 & \ding{55}\\
     HDI & 3.26 & \ding{51} & 0.02 & \ding{51}\\
     \bottomrule
    \end{tabular}
   \end{adjustbox}
\end{table}

Finally, we consider the example from the introduction, in which we want to investigate the depression rate conditioned on age, sex and place of residence. We are given the conditional means for the depression rate given age, sex, and the federal state of Germany, in addition to the joint distribution of age, sex, and state \citep{gesundheit2021depression,gesundheit2021sexAge}. Using this information, we can find the MAXENT solution for the joint distribution of all four variables (depression rate ($D$), age ($A$), sex ($S$), and place of residence ($P$)). The found Lagrange multipliers are shown in \cref{sec:add_results}. For none of the three potential causes, the multipliers are constant. Hence, we assume that all three factors (age, sex, and place of residence) have a direct causal link to the depression rate.\footnote{Note, it is still possible that these factors only have an {\it indirect} influence on the depression rate via other factors that we do not consider here. Investigating {\it all} potential causes for the depression rate would be a research project of its right and is out of the scope of this work.} Nevertheless, we can use the result from \cref{th:predictive_power}, stating that the joint MAXENT solution is a better predictor than any of the given marginal distributions, to investigate questions like \enquote{What is the probability for a 30-year-old woman living in a certain federal state to become depressed?}. When we, for instance, consider the federal states Baden-Wuerttemberg (BW) and Berlin (BE), then the result of the MAXENT solution is 
\begin{alignat*}{2}
&p(D\mid S=\text{female}, A=30, P=\text{BW}) &&= \phantom{1}9.5\%, \\
&p(D\mid S=\text{female}, A=30, P=\text{BE}) &&= 11.2\%,
\end{alignat*}
while from the marginal distributions, we get 
\begin{alignat*}{5}
&p(D\mid S=\text{female}) &&= 9.7\% , &\; &p(D\mid P=\text{BW}) &&= 7.7\% , \\
&p(D\mid A=30-44) &&= 7.5\%  , &\; &p(D\mid P=\text{BE}) &&= 9.3\%  .
\end{alignat*}

It seems surprising that the probability increases when conditioning on all three factors. However, since the depression rate for \enquote{female} is higher than for \enquote{male} (which is only 8.6\%), it makes sense that the depression probability slightly increases when additionally conditioning on the sex being female. Further note that none of the above necessarily reflects the true probability. The MAXENT solution only provides a \enquote{better guess} for the depression rate given all three factors than each of the marginal distributions.


\section{CONCLUSION}\label{sec:conclusion}

We have derived how the MAXENT principle can identify links in causal graphs and thus obtain information about the causal structure by merging the statistical information in different datasets.

There are several directions of extension of this work. On the practical side, we believe that developing efficient ways to compute the expectations of the inferred distribution is vital. In our experiments, we merged between two and five datasets and used up to 22 constraints. In order to scale the problem to more variables and constraints, the main bottleneck is the estimation of the partition function ($\alpha$ and $\beta(\bar{\bx})$ in \cref{sec:maxent}). Some efficient ways to compute this are developed in \cite{wainwright2008graphical}, however, the properties with respect to causality remain unknown.

Another direction for future work would be to study the statistical properties of the estimated parameters. In other words, to develop a statistical of the null hypothesis of a multiplier being zero (in the case of unconditional moments, or equal to others in the case of conditional moments).

In addition to the causal insights we get from this work, we would like to highlight two ways in which this work can positively impact society. First, by using only information from expectations, a characteristic that makes MAXENT a flexible approach, we move one step forward to avoid identifying individuals in adversarial attacks. Second, by using information from different sources, we can avoid being unable to answer causal questions, or worse, giving wrong causal answers because of a lack of jointly observed data.

\section*{Acknowledgements} We thank Steffen Lauritzen for helpful remarks on undirected graphical models. 

\setlength{\itemindent}{-\leftmargin}
\bibliography{merging_datasets_paper}

\newpage
\clearpage
\onecolumn
\aistatstitle{Supplementary Materials}

\appendix

\section{GRAPHICAL CAUSAL MODELS}\label{sec:graphical_models}

In graphical causal models, the causal relations among random variables are described via a directed acyclic graph (DAG) $G$, where the expression $X_i \rightarrow X_j$ means that $X_i$ influences $X_j$ 'directly' in the sense that intervening on $X_i$ changes the distribution of $X_j$ if all other nodes are adjusted to fixed values. 
If no hidden variable $U\notin\bX$ exists that causes more than one variable in $\bX$, then the set $\bX$ is said to be {\it causally sufficient} \citep{spirtes2010introduction}. 

The crucial postulate that links statistical observations with causal semantics is the {\it causal Markov condition} \citep{Spirtes1993,Pearl2000}, stating that each node $X_n$ is conditionally independent (CI) of its non-descendants given its parents $PA(X_n)$ w.r.t.\ the graph $G$. Then, the probability mass function of the joint probability distribution factorises into 
\begin{align}
p(x_1,\dots x_N) = \prod_{n=1}^N p(x_n \mid pa(x_n)) \;,
\end{align}
where $p(x_n \mid pa(x_n))$ are often called {\it Markov kernels} \citep{Lauritzen}. This entails further CIs described by the graphical criterion of {\it d-separation} \citep{Pearl2000}. 
The Markov condition is a necessary condition for a DAG being causal.
To test the corresponding CIs is a first sanity check for a causal hypothesis. 

More assumptions are required to infer causal structure from observational data. One common assumption is {\it faithfulness}: a distribution is faithful to a DAG $G$ if a CI in the data implies d-separation in the graph. Inferring the entire causal DAG from passive observations, or {\it causal graph discovery}, is, nevertheless, an ambitious task \citep{Spirtes1993, peters2017elements}. We, therefore, focus on the weaker task of inferring the presence or absence of certain causal links.

\section{DATASETS COMING FROM DIFFERENT JOINT DISTRIBUTIONS}\label{app:different_contexts}
Our approach implicitly assumes that all datasets are taken from the same joint distribution. This assumption deserves justification.
Suppose, for instance, we are interested in statistical relations between variables $X_1,\dots, X_N$ describing different health conditions of human subjects.
Assume we are given $L$ datasets containing different subsets of variables (e.g. bivariate statistics), but the datasets are from different countries.
Accordingly, we should not assume a common joint distribution $X_1,\dots, X_N$. Instead, we may introduce an additional variable $C$, and a dataset from country $C=c$ containing variables $X_i,X_j$ then provides only information about $E[f(X_i,X_j)|C=c]$. We would then infer a joint distribution of $C,X_1,\dots,X_N$ via MAXENT, given the conditional expectations.


\section{PROOFS}\label{app:proofs}
Here we repeat the theorems, corollaries and lemmas from the main text and provide the complete proofs for all of them.

\cioneway*
\begin{proof}
We first show that CI w.r.t.\ $P$ results in CI w.r.t.\ the MAXENT distribution. Let $Q$ be a distribution satisfying the following two conditions: 
\begin{itemize}
\item[(a)] $Q(X_i,X_j)=P(X_i,X_j)$, $Q(X_i,\bZ)=P(X_i,\bZ)$, and $Q(X_j,\bZ)=P(X_j,\bZ)$, and 
\item[(b)] $X_i\CI X_j\mid \bZ \;\; [Q]\;$.
\end{itemize} 
We know that such a distribution satisfying (a) and (b) exists, as this is the case for at least $P$ itself. Now assume that the MAXENT distribution $\hat{P}$ satisfies condition (a) but not condition (b). Then the entropy of $\hat{P}$ is
\begin{align*}
H_{\hat{p}}(\bX) &= H_{\hat{p}}(X_i\mid X_j,\bZ) + H_{\hat{p}}(X_j,\bZ) 
\stackrel{\centernot{\text{(b)}}}{<} H_{\hat{p}}(X_i\mid \bZ) + H_{\hat{p}}(X_j,\bZ)  \\
&\stackrel{\text{(a)}}{=}H_{q}(X_i\mid \bZ) + H_{q}(X_j,\bZ) 
 \stackrel{\text{(b)}}{=} H_{q}(X_i\mid X_j,\bZ) + H_{q}(X_j,\bZ) = H_{q}(\bX) 
    \; .
\end{align*}
This violates the assumption that $\hat{P}$ maximises the entropy. Hence, the distribution satisfying the marginal constraints in (a) that maximises the entropy must satisfy the CI in (b). 

Next, we show that CI w.r.t. the MAXENT distribution results in the respective Lagrange multipliers being zero. 
By applying Bayes' rule to the MAXENT distribution in \cref{eq:maxent_sol} it can be seen that
\begin{align*}
\hat{p}(x_i\mid x_j,\bz) &=  \frac{p(\bx)}{\sum_{x_i}p(\bx)} 
= \frac{\exp\left[\sum_{k} \lambda_k f_k(\bx_{S_k}) + \alpha\right]}{\sum_{x_i}\exp\left[\sum_{k} \lambda_k f_k(\bx_{S_k}) + \alpha\right]} \\
&=  \frac{\exp\left[\sum\limits_{{k \text{ with }}\atop{\bX_{S_k}=\{X_i\}\cup\bZ}} \lambda_k f_k(\bx_{S_k})  
 + \sum\limits_{{k \text{ with }}\atop{\bX_{S_k}=\{X_i,X_j\}}} \lambda_k f_k(\bx_{S_k}) + \alpha\right]}
{\sum\limits_{x_i}\exp\left[\sum\limits_{{k\text{ with }}\atop{\bX_{S_k}=\{X_i\}\cup\bZ}} \lambda_k f_k(\bx_{S_k}) 
+ \sum\limits_{{k \text{ with }}\atop{\bX_{S_k}=\{X_i,X_j\}}} \lambda_k f_k(\bx_{S_k}) + \alpha\right]}
\end{align*}
Using the linear independence of the functions $f$, it directly follows that 
\begin{align*}
&\hat{p}(x_i\mid x_j,\bz) = \hat{p}(x_i\mid \bz) \quad \\
\Rightarrow \quad &\lambda_k = 0  \quad\forall k \;\text{ with }\; \bX_{S_k}=\left\{X_i,X_j\right\}
\end{align*}
and from this, it directly follows the assertion.
\end{proof}

An alternative way to prove \cref{th:ci_one_way} is by considering an undirected graphical model and using insights from information geometry. 
To do this, we consider an undirected graph $G_U$ with a vertex set corresponding to the random variables $\bX$. Furthermore, let the joint distribution $P(\bX)$ satisfy the global Markov condition on $G_U$ and have strictly positive density $p(\bx)>0$. Then the Hammersley-Clifford theorem \citep{Lauritzen} tells us that the joint density $p(\bx)$ can be factorised into 
\begin{align}
p(\bx) = \frac{1}{\tilde{\alpha}} \prod_{C\in\cC} \psi_C(\bx_C) \qquad 
\end{align}
with
\begin{align}
\tilde{\alpha} = \sum_{\bx} \prod_{C\in\cC} \psi_C(\bx_C)
\end{align}
for some clique potentials $\psi_C: \cX_C\to\left[0,\infty)\right.$, where $\cC$ is the set of maximal cliques of the graph $G_U$ and $\bX_C$ are the variables corresponding to the nodes in clique $C$. 
In a log-linear model, we can formulate the clique potentials as
\begin{align}
\psi_C(\bx_C) = \exp\left[\sum_{k=1}^K \theta_{C,k} h_k(\bx_C)\right] 
\end{align}
for some measurable functions $h_k: \cX_C\to\R$. 
Hence, the joint density can be written in the form 
\begin{align}\label{eq:p_factor_HC}
p(\bx) = \exp\left[\sum_{C,k} \theta_{C,k} h_k(\bx_C) - \tilde{\alpha}\right] \; .
\end{align}
This strongly resembles the MAXENT distribution (see \cref{eq:maxent_sol}). And indeed, if the subsets of variables $\bX_{S_k}$ observed in the different datasets would be equal to the maximal cliques of the undirected graph, there would be a one-on-one correspondence between the MAXENT solution and the factorised true distribution. As a result, we would directly get equivalence in \cref{eq:ci_one_way} in \cref{th:ci_one_way}. In general, however, this is not the case. Nevertheless, the clique potential formalism provides an additional way to prove \cref{th:ci_one_way}. 

\begin{proof}[Alternative proof for \cref{th:ci_one_way}]
Let us, without loss of generality, assume that $Z=\bZ$ is one (vector-valued) variable. Then $X_i\CI X_j\mid Z$ w.r.t.\ $P$ implies that $P$ can be represented by the undirected graphical model $X_i - Z - X_j$ \citep{Lauritzen} or a subgraph of it (in case $X_i$ or $X_j$ are also independent of $Z$). Accordingly, $P$ factorises according to the clique potentials of this graph and \cref{eq:p_factor_HC}. Thus $P$ lies in the exponential manifold \citep{Amari} $\tilde{E}$ of distributions given by $\exp\left[h_1(x_i,z) + h_2(x_j,z)-\tilde{\alpha}\right]$ with arbitrary functions $h_1,h_2$. Let $E\supset \tilde{E}$ be the exponential manifold of distributions $\exp\left[h_1(x_i,z) + h_2(x_j,z)+ h_3(x_i,x_j)-\tilde{\alpha}\right]$ with arbitrary functions $h_1,h_2,h_3$. By elementary results of information geometry \citep{Amari}, $\hat{P}$ can also be defined as the projection of $P$ onto $E$. Since $P$ lies in $\tilde{E}$, and thus also in $E$, it follows that $P=\hat{P}$ in this case. This also implies that $h_3(x_i,x_j)=0$ in the MAXENT distribution and thus $\sum_k\lambda_kf_k(x_i,x_j)=0$. Due to the linear independence of the functions $f$ this implies that $\lambda_k=0$ for all $k$ with $\bX_{S_k}=\left\{X_i,X_j\right\}$. 
\end{proof}

\fexpectations*
\begin{proof} 
If $X_i$ and  $X_j$ are connected in $G$, the distribution $q(\bx)\sim \exp\left[ f_k(x_i,x_j)\right]$ is Markov relative to $G$. Obviously, it is $\lambda_k^Q=1\neq 0$, and due to $P$ having faithful $f$-expectations it is also $\lambda^P_k\neq 0$.   
\end{proof}

\cibothways*
\begin{proof}
The first statement follows from \cref{th:ci_one_way}, and the second statement directly follows from \cref{lm:f-expectations}.
\end{proof}

\noedge*
\begin{proof}
This directly follows from \cref{th:ci_both_ways}.
\end{proof}

\moral*
\begin{proof}
The undirected graph $G^b$ contains all edges of $G$ due to \cref{lm:f-expectations}. It only remains to show that $G^b$ also connects pairs with a common child. To show that $G^b$ also connects pairs $X_i,X_j$ with a common child $X_c$, we first consider the 3-node DAG $X_i \rightarrow X_c \leftarrow X_j$, and construct an example distribution, that is Markovian for this DAG, which uses only pair-interactions, including an interaction term $X_i,X_j$. By embedding this distribution into a general joint distribution, we conclude that common children can result in interaction terms after projection on pair interactions. 

We define a Markovian distribution $P$ via $P(X_i) P(X_j) P(X_c\mid X_i,X_j)$, with 
\begin{align*}
P(X_c\mid X_i,X_j) := \exp &\left[ \phi_i (X_c,X_i) + \phi_j(X_c,X_j)  - \log z(X_i,X_j)\right] \;,
\end{align*} 
where the partition function $z$ reads 
\begin{align*}
z(X_i,X_j) :=   \sum_{x_c}   \exp \left[ \phi_i (x_c,X_i) + \phi_j(x_c,X_j)\right]\;.
\end{align*} 
By construction, $P$ lies in the exponential manifold spanned by univariate and bivariate functions. It therefore coincides with the MAXENT distribution subject to all bivariate marginals. Thus, $G^b$ contains the edge $X_i - X_j$ whenever $z(X_i,X_j)$ depends on both $X_i$ and $X_j$. This dependence can be easily checked, for instance, for $\phi_i(x_c,x_i):=\delta_{x_c} \delta_{x_i}$ and  $\phi_j(x_c,x_j):=\delta_{x_c} \delta_{x_j} $, where $\delta_{x_i},\delta_{x_j},\delta_{x_c}$ are indicator functions for arbitrary values, as defined in \cref{eq:delta}.

For any DAG $G$ with $N$ variables containing the collider above as subgraph, $P(X_1,\dots,X_N) \sim P(X_i,X_j,X_c)$ is also Markov relative to $G$ and, at the same time, coincides with the MAXENT solution subject to the bivariate constraints. Hence the moral graph $G^m$ still has an edge $X_i - X_j$ because there exists a distribution, Markovian to $G$, that has a bivariate term depending on $X_i$ and $X_j$ in the MAXENT distribution subject to all bivariate distributions.
\end{proof}

\predictive*
\begin{proof}
    The proof follows directly from the duality of MAXENT and maximum likelihood \citep{wainwright2008graphical}. 
    However, we prove it here using the Lagrange multipliers found by the optimisation procedure.

    By the definition of maximum likelihood and MAXENT, we can write:
    \begin{align}\label{eq:max_likelihood}
        \Exp{P(X_j, X_i, \bZ)}{\log \hat{P}(x_j \mid x_i, \bz)} &= 
        \mathbb{E}_{P(X_j, X_i, \bZ)}\left[\log \max_{\lambda} \exp\left(\sum_{k} \lambda_{k}f_{k}(x_j, \bz) 
       + \sum_{l} \lambda_{l}g_{l}(x_j, x_i) 
      -\beta(x_i, \bz)\right)\right] 
    \end{align}
    On the other hand, if we do not use the MAXENT solution, 
    the maximum likelihood estimate we can attain consistent with $P(X_j, X_i)$ is $P(X_j \mid X_i)$.
    From \cref{eq:max_likelihood}, we can attain that solution by setting all $\lambda_{k}$ to zero.
    This means that if $P(X_j, \bZ)$ is not valuable in predicting the multipliers, then we attain the same solution as not using the information from $P(X_j, \bZ)$.
    However, if there is information to be exploited from the moments given by $P(X_j, \bZ)$, then the multipliers are not set to zero, attaining a higher likelihood.
\end{proof}


\section{OBTAINING INFORMATION ABOUT THE STRENGTH OF CAUSAL EFFECTS BY MERGING DATASETS}\label{sec:causal_influence}

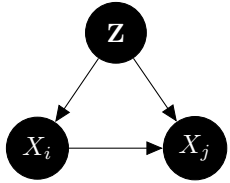
\begin{figure}[t]
\centering
\adjustbox{max width=\textwidth}{

\begin{tikzpicture}
\node[obs] (Z) {$\bZ$};
\node[below=1cm of Z] (dots) {};
\node[obs,left=.5cm of dots] (X1) {$X_i$};
\node[obs,right=.5cm of dots] (XN) {$X_j$};

\edge{Z}{X1,XN};
\edge{X1}{XN};

\end{tikzpicture}}
\caption{DAG for a treatment variable $X_i$ influencing a target $X_j$ in the presence of confounders $\bZ$. \label{fig:xyz}}
\end{figure}

Another similarly essential and challenging task is to quantify the causal influence of a treatment on a target in the presence of confounders. 
In this section, we consider a scenario where we want to investigate the causal effect of a treatment variable $X_i$ (e.g.\ the place of residence) on a target variable $X_j$ (e.g.\ the depression rate) in the presence of confounders $\bZ$ (e.g.\ the age that can influence both the depression rate and the place of residence, as displayed in \cref{fig:xyz}). Only pairwise observations for treatment -- target and treatment -- confounders are available in this scenario. To investigate the causal effect of $X_i$ on $X_j$, first, we can use the results from \cref{sec:merging_datasets} and the MAXENT distribution to identify if there is a direct causal link from $X_i$ to $X_j$. If this is the case, this section provides further insights into the causal relationship between $X_i$ and $X_j$. Even without observing all variables jointly, we can derive bounds on the interventional distribution $P(X_j\mid do(X_i))$ and the ACE of $X_i$ on $X_j$.   

\paragraph{Background}
One of the core tasks in causality is computing interventional distributions. By answering the question ``what would happen if variable $X_i$ was set to value $x_i$?" they provide valuable information without actually having to perform an experiment in which $X_i$ is set to the value $x_i$. Pearl's {\it do-calculus} \citep{Pearl2000} provides the tools to compute the distribution $P(X_j\mid do(X_i))$ of $X_j$ when intervening on $X_i$. In the infinite sample limit, the interventional distribution can be computed non-parametrically using backdoor adjustment
\begin{align}\label{eq:bd}
p(x_j\mid do(x_i)) = \sum_{\bz} p(x_j\mid x_i,\bz) p(\bz) \; ,
\end{align}
if $\bZ\subseteq \bX\backslash\left\{X_i,X_j\right\}$ is a set of nodes that contains no descendent of $X_i$ and blocks all paths from $X_i$ to $X_j$ that contain an arrow into $X_i$ \citep{Pearl2000}. 
In the case of binary variables, the interventional distribution can also be used to compute the average causal effect (ACE) of $X_i$ on $X_j$:
\begin{align}\label{eq:ACE}
ACE_{X_i\to X_j} 
= &p(x_j=1\mid do(x_i=1)) 
-p(x_j=1\mid do(x_i=0)) \; .
\end{align}

\paragraph{Deriving Bounds on the Interventional Distribution and the ACE}  Using the backdoor adjustment in \cref{eq:bd} we can bound the interventional distribution based only on the observed marginal distributions.
 
\begin{Theorem}\label{th:bounds}
Let $X_i$, $X_j$, and $\bZ$ be discrete random variables in the causal DAG shown in \cref{fig:xyz} with known marginal distributions $P(X_i,X_j)$ and $P(X_i,\bZ)$. Then the interventional distribution $P(X_j\mid do(X_i))$ is bounded as follows:
\begin{align}\label{eq:bounds}
 \frac{p(x_j,x_i={x}'_i)}{\max_{\bz} p(x_i={x}'_i\mid \bz)} 
&\leq p(x_j\mid do(x_i=x'_i)) 
\leq  \frac{p(x_j, x_i={x}'_i)}{\min_{\bz} p(x_i={x}'_i\mid \bz)}  \; .
\end{align} 
\end{Theorem}

\begin{proof}
Using Pearl's backdoor adjustment in \cref{eq:bd} and Bayes' rule, we find
\begin{align*}
p(x_j\mid do(x_i={x}'_i)) 
&= \sum_{\bz} p(x_j\mid x_i={x}'_i, \bz)p(\bz) 
= \sum_{\bz} p(x_j\mid x_i={x}'_i, \bz)p(\bz)\cdot \frac{p(x_i={x}'_i\mid \bz)}{p(x_i={x}'_i\mid \bz)} \\
&\leq \frac{\sum_{\bz} p(x_j\mid x_i={x}'_i, \bz)p(\bz)p(x_i={x}'_i\mid \bz)}{\min_{\bz} p(x_i={x}'_i\mid \bz)} 
= \frac{p(x_j, x_i={x}'_i)}{\min_{\bz} p(x_i={x}'_i\mid \bz)} \; .
\end{align*}
The lower bound can be derived analogously. 
\end{proof}

In the case where $X_i$ and $X_j$ are binary, we can use \cref{eq:bounds} also to bound the ACE of $X_i$ on $X_j$.

\begin{Lemma}\label{lm:ace}
In the setting described in \cref{th:bounds} the ACE of $X_i$ on $X_j$ is bounded as follows:
\begin{align}
 \frac{p(x_j\text{=}1, x_i\text{=}1)}{\max_\bz p(x_i\text{=}1\mid \bz)}  - \frac{p(x_j\text{=}1, x_i\text{=}0)}{\min_{\bz}p(x_i\text{=}0\mid \bz)}
\leq ACE_{X_i\to X_j} \leq \frac{p(x_j\text{=}1, x_i\text{=}1)}{\min_{\bz}p(x_i\text{=}1\mid \bz)}  - \frac{p(x_j\text{=}1, x_i\text{=}0)}{\max_{\bz}p(x_i\text{=}0\mid \bz)} \; .
\end{align}
\end{Lemma}

\begin{proof}
This directly follows from \cref{th:bounds} and \cref{eq:ACE}.
\end{proof}

\Cref{lm:ace} provides us with at least an approximate insight of the strength of the causal effect of the variable $X_i$ on $X_j$.
In addition, the bounds provide a correct scale of the bounds that could be found using the MAXENT solution.
However, as the MAXENT solution is an approximation to the true distribution, this will also be an approximation, and as it is a point estimate, we do not know how close or far we are from the true ACE. 
Therefore we report bounds to show what can be said from marginal distributions about the ACE even without MAXENT.  
 
 \paragraph{Related Work on Confounder Correction} The classical task of confounder correction is to estimate the effect of a treatment variable on a target in the presence of unobserved confounders. In this paper, however, we consider the scenario shown in \cref{fig:xyz} and assume that we have observations for the confounders $\bZ$, but not for $X_i,X_j$, and $\bZ$ jointly. 
If $X_i,X_j$, and $\bZ$ were observed jointly, the causal effect of $X_i$ on $X_j$ would be identifiable and could be computed using Pearl's backdoor adjustment (see \cref{sec:graphical_models}). In cases where $\bZ$ is unobserved, the causal effect of $X_i$ on $X_j$ is not directly identifiable. One exception is if a set of observed variables satisfies the front-door criterion \citep{Pearl2000}. In \citet{Galles1995,Pearl2000} and \citet{kuroki2014measurement} more general conditions were presented for which do-calculus and proxy-variables of unobserved confounders, respectively, make the causal effect identifiable. 
Another approach to confounder correction is by phrasing the problem in the potential outcome framework, e.g., using instrumental variables \citep{angrist1996identification,GroJanSieSch2016} or principal stratification \citep{Rubin2004}. 
Other, more recent approaches include, for instance: double/debiased machine learning \citep{chernozhukov2018double,jung2021estimating}; combinations of unsupervised learning and predictive model checking to perform causal inference in multiple-cause settings \citep{wang2019blessings}; methods that use limited experimental data to correct for hidden confounders in causal effect models \citep{kallus2018removing}; and the split-door criterion which considers time series data where the target variable can be split into two parts of which one is only influenced by the confounders and the other is influenced by the confounders and the treatment, reducing the identification problem to that of testing for independence among observed variables \citep{sharma2018split}. 
Although confounder correction is a common and well-studied problem, we are unaware of approaches based on pairwise observations for treatment -- target and treatment -- confounders only.


\section{ADDITIONAL RELATED WORK}\label{app:add_related_work}

\paragraph{Gaining statistical information from causal knowledge}
One approach to using causal information to improve the approximation to the true joint distribution is {\it causal} MAXENT \citep{sun2006causal,janzing2009distinguishing}, a particular case of conditional MAXENT, where the entropy of the variables is maximised along the causal order. 
For cause-effect relations, it just amounts to first maximising the entropy of the cause subject to all constraints that refer to it. 
Then, it maximises the conditional entropy for the effect given the cause subject to all constraints.
Maximising the entropy in the causal order results in a distribution with lower entropy than maximising the entropy jointly.
Consequently, the distribution learned in the causal order will have better predictive power.
Another simple example of how causal information can help gaining statistical insights is the following: 
Imagine we are given the bivariate marginal distributions $P(X_1,X_2)$ and $P(X_2,X_3)$. 
In the general case, where we do not know the causal graph, we could not identify the joint distribution. 
However, when we know that the three variables form a causal chain $X_1\to X_2\to X_3$, this causal information is enough to identify the joint distribution uniquely \citep{overlapping}.
\footnote{In general, it is a non-trivial problem to decide whether a set of marginal distributions of different but non-disjoint sets of variables are consistent with a joint distribution (the so-called \enquote{marginal problem} \citep{Vorobev1962}).} 
Even perfect causal knowledge does not uniquely determine the joint distribution for less simplistic scenarios. 
But causal information may still help to get {\it some} properties of the joint distribution. 
In \citet{Tsamardinos}, for instance, the causal structure is used to predict CI of variables that have not been observed together. 
This paper approaches a complementary problem: gaining causal insights by merging statistical information from different datasets. 

\paragraph{Entropy based approaches to extract or exploit causal information} 
Different methods exposing the relationship between information theory and causality are present in the literature. 
In \citet{kocaoglu2017entropic,compton2021entropic} properties of the entropy are used to infer the causal direction between categorical variable pairs. 
Their main idea is that if the entropy of the exogenous noise of a functional assignment in a structural causal model (SCM) is low, then the causal direction often becomes identifiable. 
Their approach differs from ours in several respects: first, we investigate the absence and presence of causal edges from merged data, as opposed to trying to infer the causal direction; 
second, we are not constrained to variable pairs; finally, we use the entropy as the function we want to optimise directly while they compare the entropy of each of the noise variables to decide the causal direction. 
In \citet{ziebart2010modeling,ziebart2013principle} the maximum causal entropy is introduced to solve inverse reinforcement learning problems. 
Their approach is based on having knowledge about a possible causal graph, making the MAXENT computation cheaper by exploiting the causal structure of the data. 
Their work differs from ours on the type of insights we get from the MAXENT estimation: while they are trying to save computation, we are trying to identify causal edges from the Lagrange multipliers.

\paragraph{Semi-supervised learning (SSL)}The relation to semi-supervised learning (SSL) is interesting but still unexplored. The high-level connection, which we can mention, is that SSL uses $P(X)$ to infer properties of $P(X,Y)$, which has been claimed to be only possible if $Y$ is the cause and $X$ the effect, but not vice versa \cite{anticausalSSL}. Hence, SSL also infers joint properties from the marginal but relies on model assumptions like cluster assumption, manifold assumption, smoothness of decision boundaries. To relate this inductive bias with MAXENT probably requires defining the correct type of functions $f$.

\section{ADDITIONAL RESULTS}\label{sec:add_results}
In this section, we provide exemplary plots of the Lagrange multipliers for the synthetic experiments discussed in \cref{sec:experiments}, as well as further results for the experiments on the two real-world datasets.  

\paragraph{Synthetic Data}

\begin{figure}[t]
\sbox0{\begin{subfigure}[t]{.3\textwidth}
    \centering
    \adjustbox{max width=.9\textwidth}{

\begin{tikzpicture}
\node[latent] (U) {$U_1$};
\node[obs,below=1cm of U] (X1) {$X_1$};
\node[obs,right=.5cm of X1] (X2) {$X_2$};
\node[obs,right=.5cm of X2] (X3) {$X_3$};
\node[obs,right=.5cm of X3] (X4) {$X_4$};
\node[obs,right=.5cm of X4] (X5) {$X_5$};
\node[latent,above=1cm of X4] (V) {$U_2$};
\node[obs,below=1cm of X3] (X0) {$X_0$};

\edge{U}{X1,X2};
\edge{V}{X3,X4};
\edge[]{X1}{X0};
\edge[]{X2}{X0};
\edge[]{X3}{X0};
\edge[]{X5}{X0};

\end{tikzpicture}}
    \subcaption{Exemplary graph (a)\label{fig:graph_exp_a_example}}
\end{subfigure}}
\sbox1{\begin{subfigure}[t]{.3\textwidth}
    \centering
    \adjustbox{max width=.9\textwidth}{

\begin{tikzpicture}
\node[latent] (U1) {$U_1$};
\node[latent,right=.5cm of U1] (U2) {$U_2$};
\node[latent,right=.5cm of U2] (U3) {$U_3$};
\node[latent,right=.5cm of U3] (U4) {$U_4$};
\node[latent,right=.5cm of U4] (U5) {$U_5$};
\node[obs,below=1cm of U1] (X1) {$X_1$};
\node[obs,below=1cm of U2] (X2) {$X_2$};
\node[obs,below=1cm of U3] (X3) {$X_3$};
\node[obs,below=1cm of U4] (X4) {$X_4$};
\node[obs,below=1cm of U5] (X5) {$X_5$};
\node[obs,below=1cm of X3] (X0) {$X_0$};

\edge{U1}{X1, X2, X3};
\edge{U2}{X2, X3, X4};
\edge{U3}{X3, X4, X5};
\edge{U4}{X4, X5, X1};
\edge{U5}{X5, X1, X2};
\edge[]{X2}{X0};
\edge[]{X3}{X0};
\edge[]{X4}{X0};
\edge[]{X5}{X0};

\end{tikzpicture}}
    \subcaption{Exemplary graph (b)\label{fig:graph_exp_b_example}}
\end{subfigure}}
\sbox2{\begin{subfigure}[t]{.3\textwidth}
    \centering
    \adjustbox{max width=.9\textwidth}{

\begin{tikzpicture}
\node[latent] (U) {$U_1$};
\node[obs,below=1cm of U] (X3) {$X_3$};
\node[obs,right=.5cm of X3] (X4) {$X_4$};
\node[obs,left=.5cm of X3] (X2) {$X_2$};
\node[obs,left=.5cm of X2] (X1) {$X_1$};
\node[obs,right=.5cm of X4] (X5) {$X_5$};
\node[obs,below=1cm of X3] (X0) {$X_0$};

\edge{U}{X1, X2, X3, X4, X5};
\edge{X1}{X2, X3};
\edge{X5}{X3, X4};
\edge[]{X5}{X0};

\end{tikzpicture}}
    \subcaption{Exemplary graph (c)\label{fig:graph_exp_c_example}}
\end{subfigure}}
\sbox3{\begin{subfigure}[t]{.3\textwidth}
    \centering
    \includegraphics[width=\textwidth]{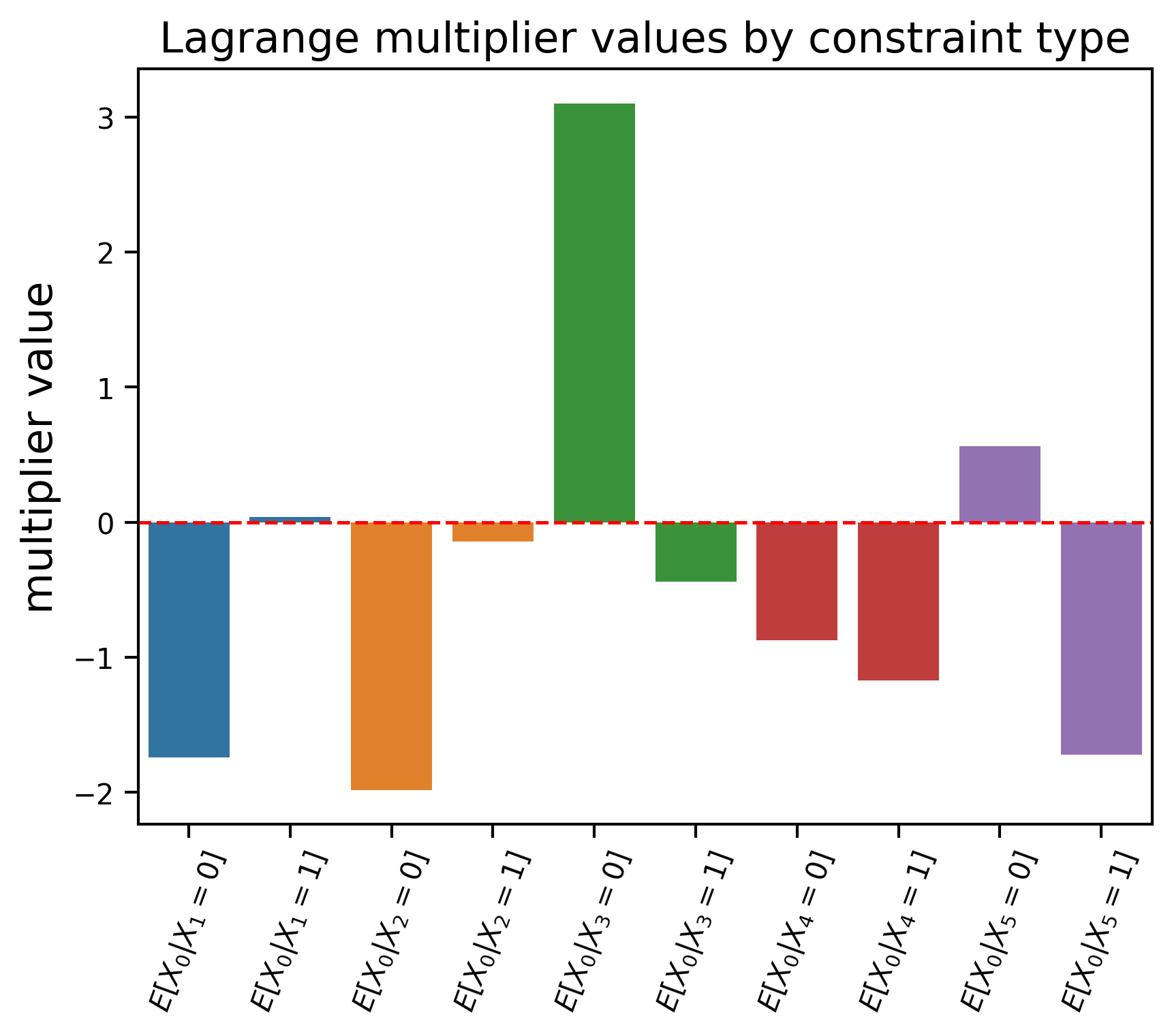}
    \subcaption{Exemplary result for the Lagrange multipliers for graph (a)\label{fig:first_multipliers}}
\end{subfigure}}
\sbox4{\begin{subfigure}[t]{.3\textwidth}
    \centering
    \includegraphics[width=\textwidth]{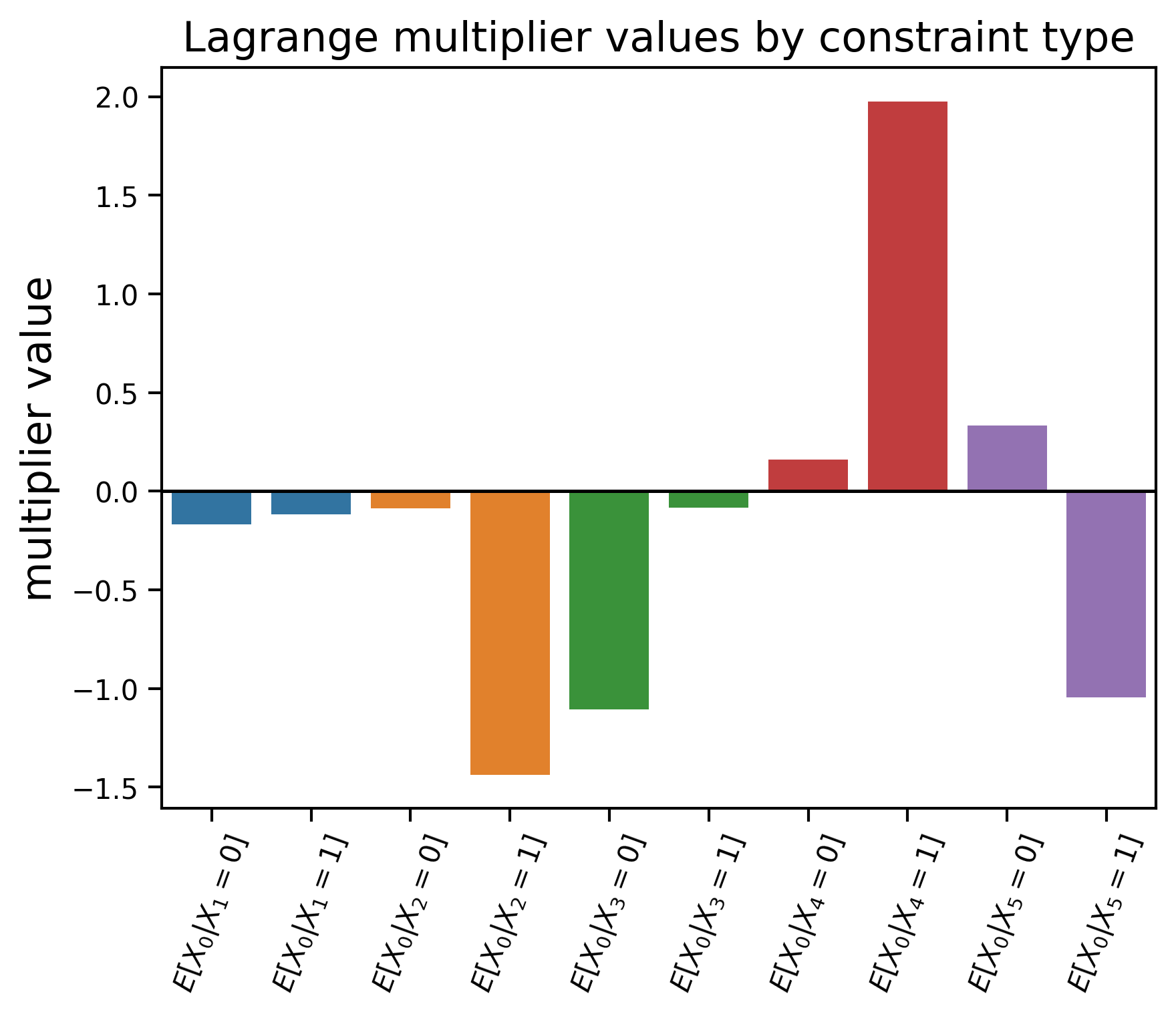}
    \subcaption{Exemplary result for the Lagrange multipliers for graph (b)\label{fig:second_multipliers}}
\end{subfigure}}
\sbox5{\begin{subfigure}[t]{.3\textwidth}
    \centering
    \includegraphics[width=\textwidth]{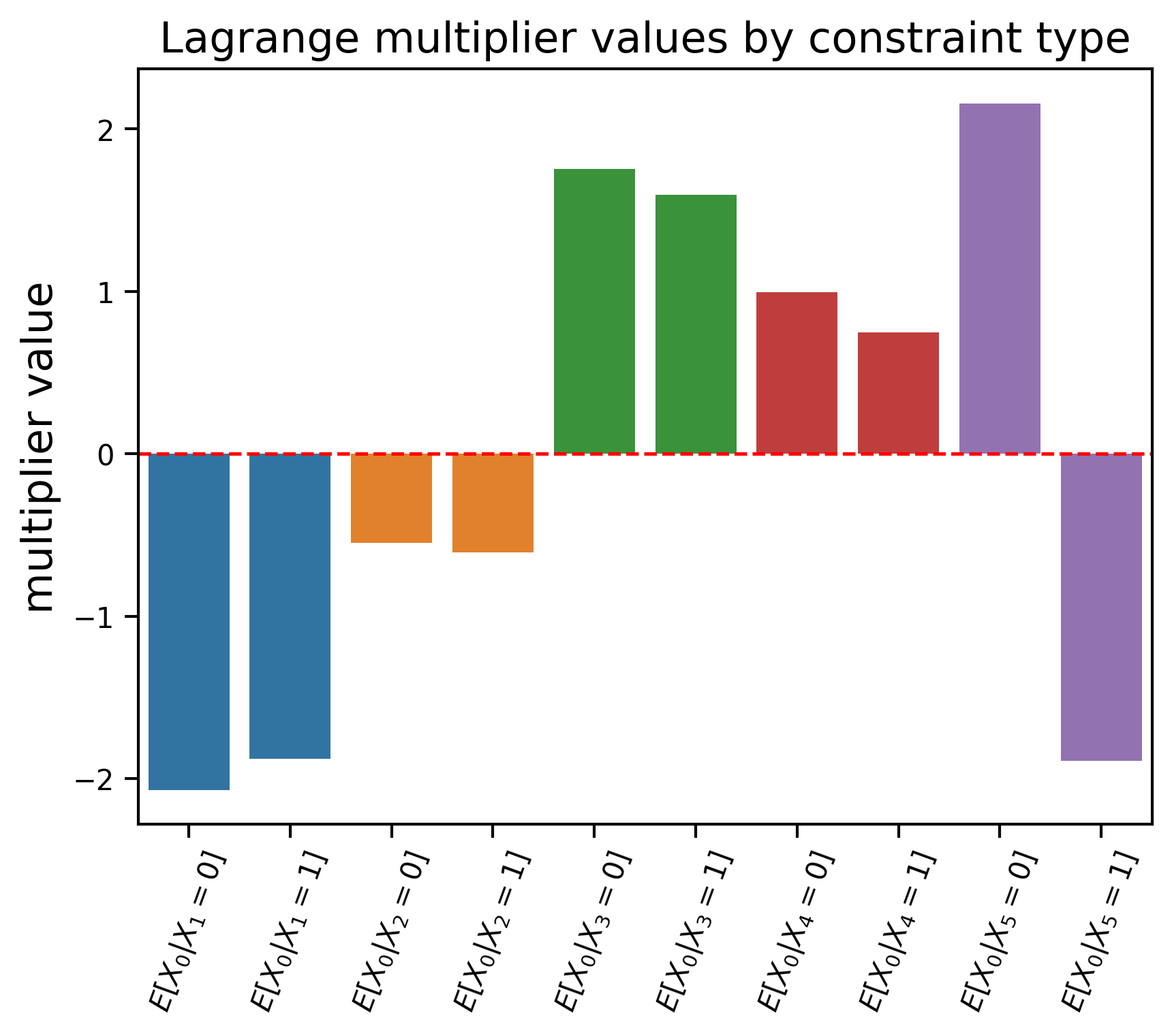}
    \subcaption{Exemplary result for the Lagrange multipliers for graph (c)\label{fig:third_multipliers}}
\end{subfigure}}
\centering
\begin{tabular}{ccc}
\usebox0  & \usebox1 & \usebox2 \\[2em]
\usebox3  & \usebox4 & \usebox5
\end{tabular}
\caption{We show the found Lagrange multipliers for a randomly picked dataset for the displayed graphs. One can see that in all three cases, the multipliers are very close to being constant whenever an edge is missing. On the other hand, the differences between them are significant in the cases where there is an edge from $X_i$ to $X_0$.\label{fig:synth_multipliers}}
\end{figure}
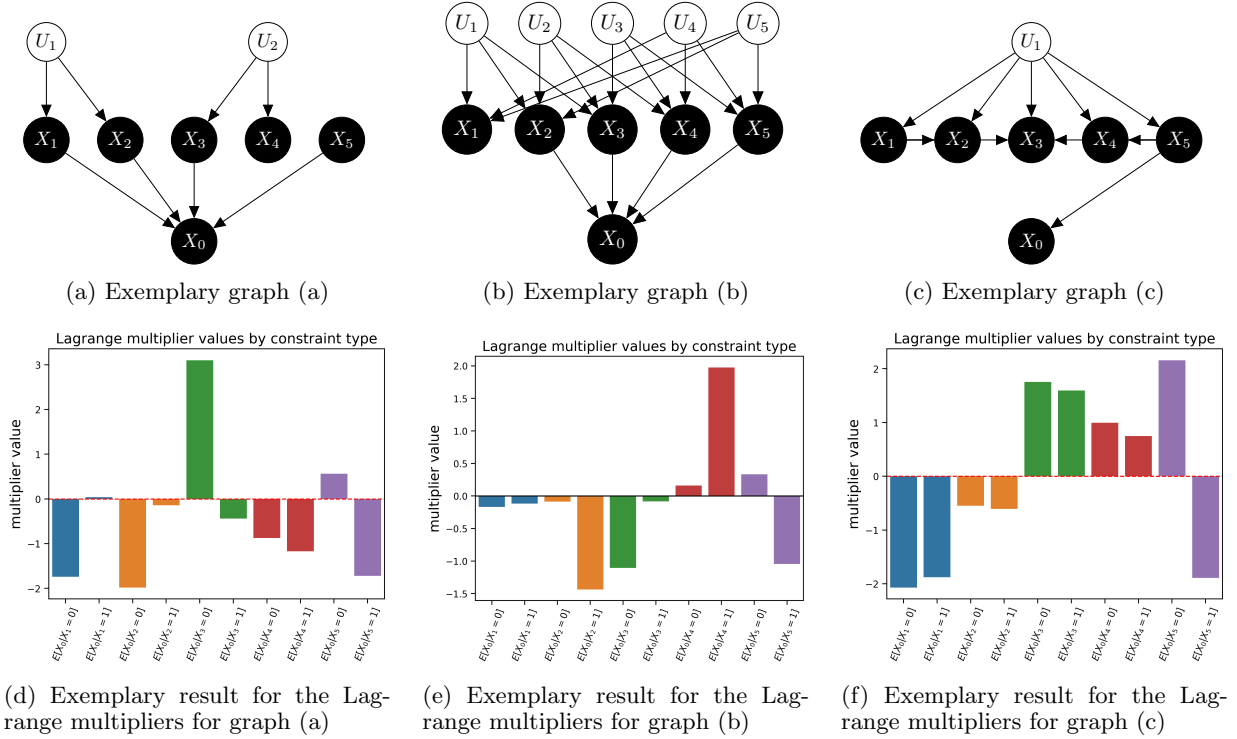

In \cref{fig:synth_multipliers} we show exemplary results for the Lagrange multipliers for the experiments with synthetic data discussed in \cref{sec:experiments}. For each of the three graphs in \cref{fig:graph_exp_a,fig:graph_exp_b,fig:graph_exp_c} we randomly picked one dataset, for which we show the exact used graph structure in \cref{fig:graph_exp_a_example,fig:graph_exp_b_example,fig:graph_exp_c_example} and the found Lagrange multipliers in \cref{fig:first_multipliers,fig:second_multipliers,fig:third_multipliers}. In all three cases, the difference between the multiplier associated with $\Exp{}{X_0\mid X_i=0}$ and the one associated with $\Exp{}{X_0\mid X_i=1}$ is very small whenever there is no edge from $X_i$ to $X_0$, and relatively large whenever there is an edge connecting $X_i$ and $X_0$.

\paragraph{Real Data}
First, we further investigate the results from the experiment on the data from \cite{Gapminder}. For this, we consider different subsets of the variables 
\begin{itemize}
\item children per woman / fertility (FER) \citep{Gapminder}; 
\item inflation-adjusted Gross Domestic Product (GDP) per capita \citep{WB:GDPPC}; 
\item Human Development Index (HDI) \citep{UNHDR:HDI}; and 
\item life expectancy (LE) in years \citep{Gapminder}.
\end{itemize} 
as potential causes of the target variable CO2 tonnes emission per capita \citep{osti_1389331}.
We always use data from 2017 for all variables and standardise it before estimation. We always use the unconditional mean and variance of CO2 emissions, and the pairwise covariance between CO2 emissions and each considered variable as constraints. We compare our results with the output of the KCI-test, where we use a significance threshold of $\alpha=0.05$. 

\begin{figure}[t]
\centering
\begin{subfigure}{.3\textwidth}
\centering
\includegraphics[width=\columnwidth]{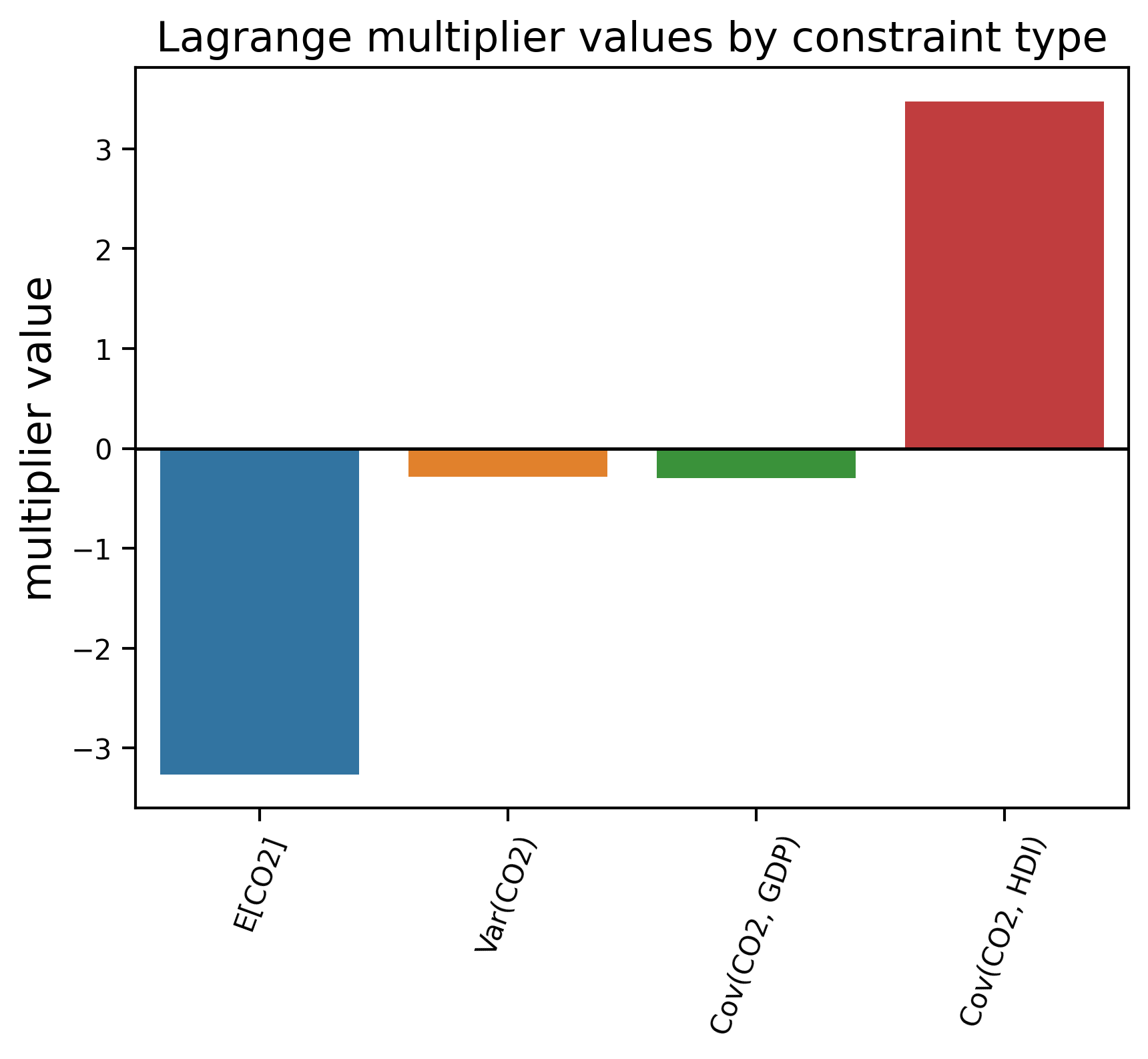}
\caption{considering GDP and HDI \label{fig:co2_gdp_hdi}}
\end{subfigure}
\hfill
\begin{subfigure}{.3\textwidth}
\centering
\includegraphics[width=\textwidth]{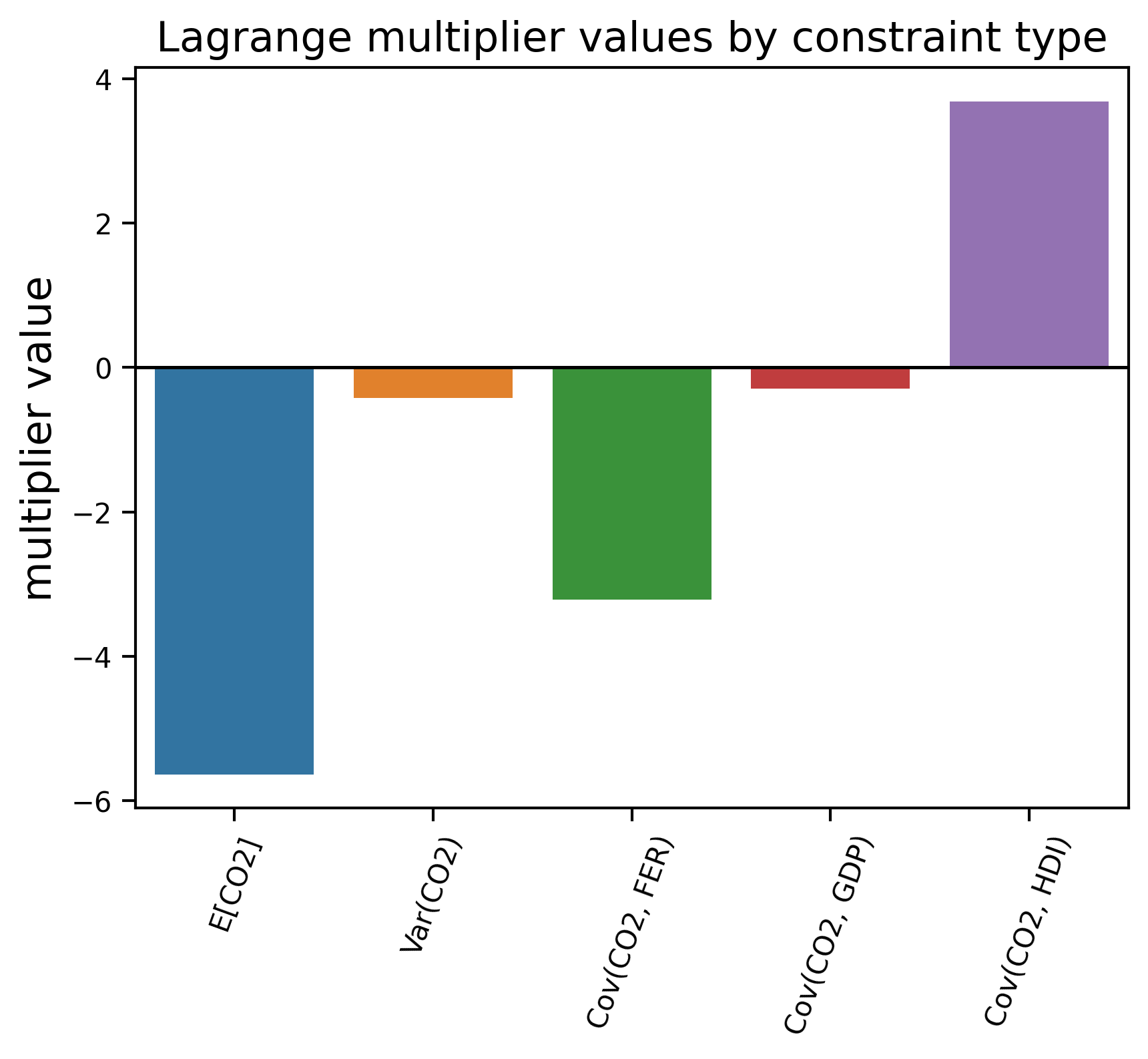}
\caption{considering FER, GDP, and HDI \label{fig:co2_fer_gdp_hdi}}
\end{subfigure}
\hfill
\begin{subfigure}{.3\textwidth}
\centering
\includegraphics[width=\textwidth]{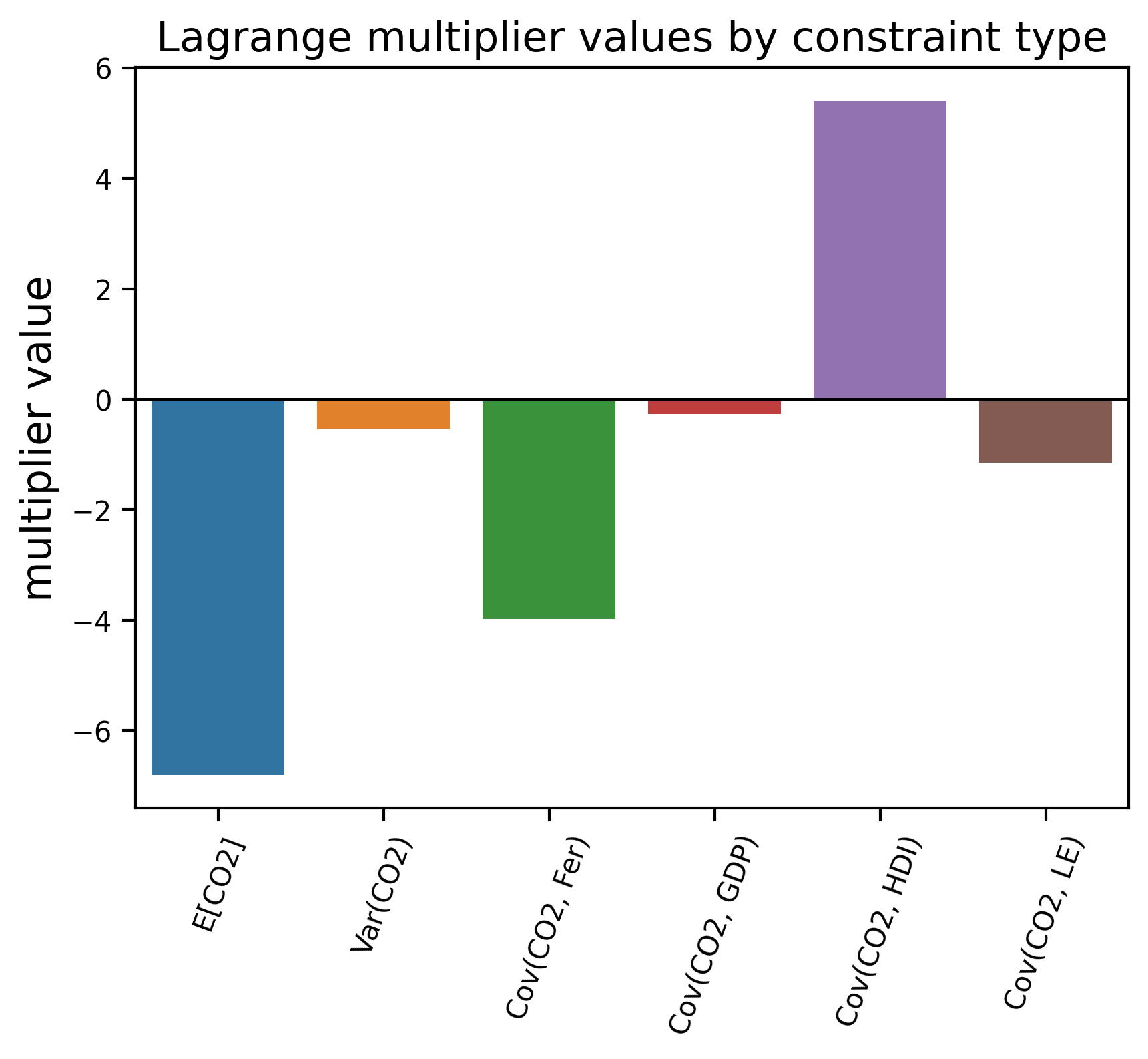}
\caption{considering FER, GDP, HDI, and LE \label{fig:co2_fer_gdp_hdi_le}}
\end{subfigure}
\caption{\label{fig:exp_results_gapminder} Lagrange multipliers for the real-world dataset from \citet{Gapminder} when considering different sets of variables as potential causes for CO2 emissions. We see that in all three cases, the results for GDP and HDI are consistent, indicating that HDI is directly linked to CO2 and GDP not.}
\end{figure}

The results in \cref{fig:exp_results_gapminder} and \cref{tab:gapminder_compare} show that the conclusions drawn from the Lagrange multipliers are consistent over the different sets of considered potential causes. For the KCI-test, on the other hand, we get separate statements about the CIs of CO2 and HDI, and CO2 and FER, depending on the considered conditioning set. At first glance, this might be not surprising as, of course, the CI relationships can change when considering more variables. For instance, one could imagine that the causal effect of HDI on CO2 is only via FER. This would explain the behaviour of the KCI-test w.r.t.\ HDI. To check if this is the case -- which would contradict the result from the Lagrange multipliers --, we perform another KCI-test for CO2 and HDI conditioned only on FER. Summarising the obtained CI statements, we get:
\begin{alignat}{2}
CO2 &\nCI HDI &&\mid GDP \label{eq:CIfirst} \\ 
CO2 &\nCI HDI && \mid FER \label{eq:CIsecond}\\
CO2 &\CI HDI &&\mid GDP, FER  \label{eq:CIthird}\\
CO2 &\CI GDP &&\mid HDI \label{eq:CIforth}\\ 
CO2 &\CI GDP && \mid HDI, FER \label{eq:CIfifth}\\
CO2 &\nCI FER &&\mid HDI, GDP \label{eq:CIsixth}
\end{alignat}
If we now try to draw a causal DAG for these four variables based on the CIs in \cref{eq:CIfirst,eq:CIsecond,eq:CIthird,eq:CIforth,eq:CIfifth,eq:CIsixth}, we see that this is not so simple. In fact, it is not possible to construct a DAG over these four variables that is consistent with \cref{eq:CIfirst,eq:CIsecond,eq:CIthird,eq:CIforth,eq:CIfifth,eq:CIsixth}. There are, of course, many possible reasons why the KCI-test provides these seemingly inconsistent results. For instance, one could argue that choosing $\alpha=0.05$ as a threshold for the decision is very arbitrary and maybe a non-optimal choice. All obtained p-values were relatively small (less than 0.2), which might indicate that the question \enquote{CI or no CI?} might not be so easy to answer in this case. Furthermore, we do not know whether the considered example violates some of the assumptions made in the KCI-test. 

This shows that the KCI-test should not be mistaken with \enquote{ground truth}, and the fact that the conclusions drawn from the Lagrange multipliers do not always coincide with the conclusions drawn from the KCI-test is not necessarily a problem of our proposed approach.

\begin{table}[t]
\centering
\caption{We show the found Lagrange multipliers $\lambda_i$ for the MAXENT solution (see (\subref{tab:co2_gdp_hdi}) to \subref{tab:co2_fer_gdp_hdi_le})) together with the p-values for the KCI-test (see (\subref{tab:co2_gdp_hdi_p}) to \subref{tab:co2_fer_gdp_hdi_le_p})) . We indicate where the multipliers and p-values indicate the presence of a direct edge connecting $X_i$ and CO2 emissions, or, respectively, that the two are not CI given the other considered variable(s). We see that the conclusions drawn from the Lagrange multipliers are consistent across the different considered sets of potential causes, while the CI statements of the KCI-test change when changing the conditioning set. \label{tab:gapminder_compare}}
   \begin{subtable}[b]{.3\textwidth}
   \centering
    \caption{considering GDP and HDI as potential causes\label{tab:co2_gdp_hdi}}
   \begin{adjustbox}{max width=\textwidth}
   \begin{tabular}{lrc}
    \toprule
    variable $X_i$ & $\lambda_i$ & edge  \\
    \midrule
    &&\\
     GDP & -0.29 & \ding{55} \\
     HDI & 3.26 & \ding{51} \\
     &&\\
    \bottomrule
    \end{tabular}
    \end{adjustbox}
    \end{subtable}
\hfill
    \begin{subtable}[b]{.3\textwidth}
    \centering
     \caption{considering FER, GDP, and HDI as potential causes\label{tab:co2_fer_gdp_hdi}}
   \begin{adjustbox}{max width=\textwidth}
    \begin{tabular}{lrc}
    \toprule
    variable $X_i$ & $\lambda_i$ & edge  \\
    \midrule
     FER  & -3.22 & \ding{51} \\
     GDP & -0.29 & \ding{55} \\
     HDI & 3.69 & \ding{51} \\
       && \\
    \bottomrule
    \end{tabular}
    \end{adjustbox}
   \end{subtable}
\hfill
    \begin{subtable}[b]{.3\textwidth}
    \centering
     \caption{considering FER, GDP, HDI, and LE as potential causes \label{tab:co2_fer_gdp_hdi_le}}
   \begin{adjustbox}{max width=\textwidth}
    \begin{tabular}{lrc}
    \toprule
    variable $X_i$ & $\lambda_i$ & edge  \\
    \midrule
     FER  & -3.98 & \ding{51}  \\
     GDP & -0.27 & \ding{55} \\
     HDI & 5.40 & \ding{51} \\
     LE & -1.15 & \ding{51} \\
    \bottomrule
    \end{tabular}
    \end{adjustbox}
   \end{subtable}
\\[2em]
   \begin{subtable}[b]{.3\textwidth}
   \centering
   \caption{considering GDP and HDI as potential causes\label{tab:co2_gdp_hdi_p}}
    \begin{adjustbox}{max width=\textwidth}
   \begin{tabular}{lrc}
    \toprule
    variable $X_i$ & p-value & no CI \\
    \midrule
    &&\\
     GDP  & 0.19 & \ding{55}\\
     HDI & 0.02 & \ding{51}\\
     &&\\
    \bottomrule
    \end{tabular}
    \end{adjustbox}
    \end{subtable}
\hfill
    \begin{subtable}[b]{.3\textwidth}
    \centering
     \caption{considering FER, GDP, and HDI as potential causes\label{tab:co2_fer_gdp_hdi_p}}
   \begin{adjustbox}{max width=\textwidth}
    \begin{tabular}{lrc}
    \toprule
    variable $X_i$  & p-value & no CI \\
    \midrule
     FER   & 0.03 & \ding{51} \\
     GDP & 0.13 & \ding{55}\\
     HDI & 0.13 & \ding{55}\\
      & & \\
    \bottomrule
    \end{tabular}
    \end{adjustbox}
   \end{subtable}
\hfill
    \begin{subtable}[b]{.3\textwidth}
    \centering
    \caption{considering FER, GDP, HDI, and LE as potential causes\label{tab:co2_fer_gdp_hdi_le_p}}
   \begin{adjustbox}{max width=\textwidth}
    \begin{tabular}{lrc}
    \toprule
    variable $X_i$  & p-value & no CI \\
    \midrule
     FER   & 0.08 & \ding{55} \\
     GDP  & 0.16 & \ding{55}\\
     HDI & 0.14 & \ding{55}\\
     LE & 0.00 & \ding{51}\\
    \bottomrule
    \end{tabular}
    \end{adjustbox}
    \end{subtable}
\end{table}

\begin{figure}[t]
\centering
\includegraphics[width=.7\textwidth]{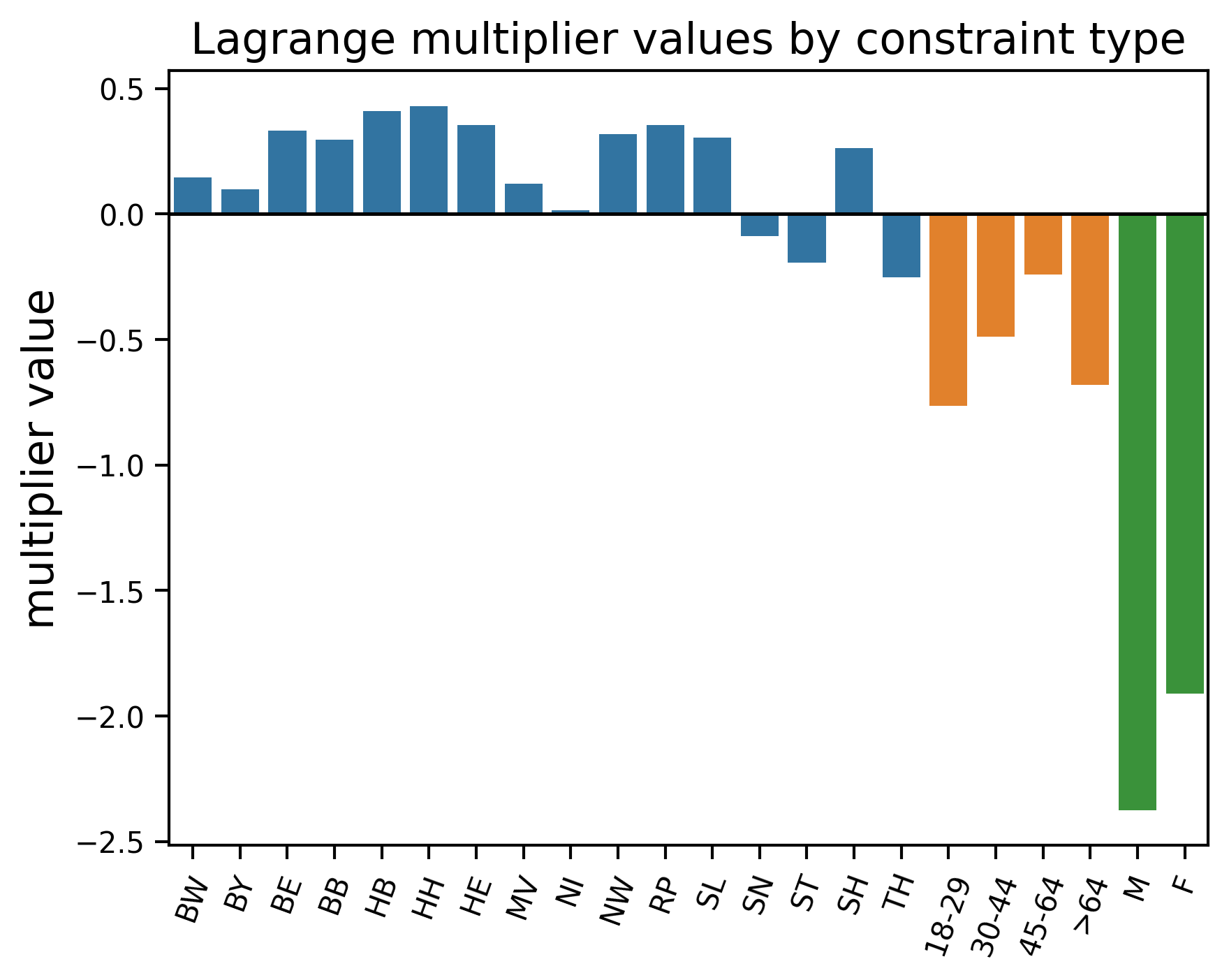}
\caption{\label{fig:exp_results_depression} Lagrange multipliers for the depression dataset. We see that for none of the three potential causes place of residence, age, and sex the multipliers are close to being constant. Hence we conclude that all three factors can directly have impact on the depression rate.}
\end{figure}

Finally, we show in \cref{fig:exp_results_depression} the Lagrange multipliers for the experiment on depression rate w.r.t.\ place of residence, age, and sex. We see that for all three factors the multipliers are {\it not} constant across the various conditions. This indicates that all three factors might directly cause the depression rate. 

\section{IMPLEMENTATION DETAILS}\label{sec:implementation}

We implemented MAXENT on Python using JAX \citep{jax2018github} optimisation procedures.
We minimise the sum of the squares of the difference between the moments given as constraints and the moments estimated using the MAXENT distribution entailed by the Lagrange multipliers.
If the absolute difference between the data expectations and the MAXENT expectations were smaller than 0.001, 
the procedure was considered convergent. 
In our current implementation, 
we estimate the normalising constant,
although there is the possibility to use approximation methods to make the computation faster, 
if required \citep{wainwright2008graphical}.

\section{EXPERIMENTAL SETUP}\label{sec:experimental_setup}

In all experiments, we observe only expectations associated with the $X_{i}$ variables. 
To build the ROC curves for each of the samples obtained, we first generated a vector $p$ of probabilities from a $\mathcal{U}(0.1, 0.9)$ distribution. 
In all the following examples, we generated 1000 observations for 100 repetitions of the SCM and estimated the empirical expectations from that sample. 
If the procedure did not converge, we did not take it into account for the ROC.
We also randomised the logical relation between the causes $X_{i}$ and the effect $X_0$.
We denote this logical relation below as $\odot \in \{\wedge, \vee, \oplus\}$.
The generative processes for the shown experiments with synthetically generated data are the following:

First, we select the used parameters as follows: 
\begin{alignat*}{2}
u_l  &\sim \cN(0, 1) \quad &&\text{for}\quad l\in\left\{1, \dots, 5\right\}\\
p_{k} &\sim \mathcal{U}(0.1, 0.9) \quad &&\text{for} \quad k=0,\dots,5 \\
a_{i} &\sim \mathcal{N}(0, 1) \quad &&\text{for} \quad i=1,\dots,5 \\
b_{i,j} &\sim \mathcal{N}(0, 1) \quad &&\text{for} \quad j=1,\dots,5 
\end{alignat*}

Then we use these parameters to generate the data for the variables $X_0$ to $X_5$. 

For the experiment in \cref{fig:graph_exp_a} the data is generated according to:
\begin{align*}
x_{1}  &\sim |\text{Ber}(p_{1}) - (u_1 > 0)| \\
x_{2}  &\sim |\text{Ber}(p_{2}) - (u_1 < 0.25)| \\
x_{3}  &\sim |\text{Ber}(p_{3}) - (u_2 > 0)| \\
x_{4}  &\sim |\text{Ber}(p_{4}) - (u_2 > 0.25)| \\
x_{5}  &\sim \text{Ber}(p_{5}) \\
x_0 &\sim \mathbf{1}_{>0} \bigg[\bigg(\sum_{i}a_{i}X_{i}+\sum_{i,j}b_{i,j}X_{i}X_{j}\bigg)\bigg] \odot \text{Ber}(p_{0}) 
\end{align*}

And finally, for the experiment in \cref{fig:graph_exp_b} the data is generated according to:
\begin{align*}
x_{1}  &\sim |\text{Ber}(p_{1}) - (u_1 > 0 \vee u_2 > 0.25 \vee u_3 > 0.5)| \\
x_{2}  &\sim |\text{Ber}(p_{2}) - (u_2 < 0.5 \vee u_3 < 0.25 \vee u_4 < 0)| \\
x_{3}  &\sim |\text{Ber}(p_{3}) - (u_3 > 0 \vee u_4 < 0.25 \vee u_5 > 0.5)| \\
x_{4}  &\sim |\text{Ber}(p_{4}) - (u_4 < 0.5 \vee u_5 > 0.25 \vee u_1 < 0)| \\
x_{5}  &\sim |\text{Ber}(p_{5}) - (u_5 > 0 \vee u_1 < 0.25 \vee u_2 > 0.5)| \\
x_0 &\sim \mathbf{1}_{>0} \bigg[\bigg(\sum_{i}a_{i}X_{i}+\sum_{i,j}b_{i,j}X_{i}X_{j}\bigg)\bigg] \odot \text{Ber}(p_{0}) 
\end{align*}

For the experiment in \cref{fig:graph_exp_c} we used the following generative process:
\begin{align*}
x_{1}  &\sim |\text{Ber}(p_{1}) - (u_1 > 0)| \\
x_{5}  &\sim |\text{Ber}(p_{5}) - (-0.25 < u_1 < 0.25)| \\
x_{2}  &\sim |\text{Ber}(p_{2}) - (u_1 < 0)| \vee x_{1}\\
x_{4}  &\sim |\text{Ber}(p_{4}) - (u_1 < -0.25)| \vee x_{5}\\
x_{3}  &\sim |\text{Ber}(p_{3}) - (u_1 > 0.25)| \vee (x_{1} \oplus x_{5})\\
x_0 &\sim \mathbf{1}_{>0} \bigg[\bigg(\sum_{i}a_{i}X_{i}+\sum_{i,j}b_{i,j}X_{i}X_{j}\bigg)\bigg] \odot \text{Ber}(p_{0}) 
\end{align*}

\end{document}